\documentclass[11pt,sort&compress]{elsarticle}
\usepackage[paper=letterpaper,top=24mm, bottom=26mm, left=26mm, right=26mm]{geometry}
\makeatletter
\def\ps@pprintTitle{
 \let\@oddhead\@empty
 \let\@evenhead\@empty
 \def\@oddfoot{\centerline{\thepage}}
 \let\@evenfoot\@oddfoot}
\makeatother

\usepackage[hyperref]{xcolor}
\definecolor{darkgreen}{rgb}{0.01, 0.75, 0.24}
\definecolor{darkblue}{HTML}{2B66D3}
\usepackage[colorlinks=true,
            linkcolor=darkblue,
            urlcolor=darkblue,
            citecolor=darkblue,linkcolor=darkblue,hyperfootnotes=true]{hyperref}
\usepackage{amsmath}
\usepackage{amsfonts}
\usepackage{accents}
\usepackage{amssymb}
\usepackage{mathrsfs}
\usepackage{mathtools}
\usepackage[utf8]{inputenc}
\usepackage[T1]{fontenc}
\usepackage{subfigure}
\usepackage{wasysym}

\usepackage[cal=boondox,calscaled=1]{mathalfa}
\DeclareMathAlphabet{\bbvar}{U}{BOONDOX-ds}{m}{n}

\usepackage{psfrag}
\usepackage{pstool}
\usepackage{caption}
\usepackage{tabularx}
\usepackage{multirow}
\usepackage{pbox}
\usepackage{graphicx}
\usepackage{csquotes}
\newcommand{\hook}{\text{\large{$\lrcorner$}}}

\newcommand{\utilde}[1]{\underaccent{\tilde}{#1}}
\newcommand{\di}{\mathrm{d}}
\usepackage{tensor}
\newcommand{\ou}[3]{\tensor{#1}{^{#2}_{#3}}}
\newcommand{\uo}[3]{\tensor{#1}{_{#2}^{#3}}}

\newcommand{\nT}[3]{\tensor[^{(#1)}]{#2}{#3}}
\newcommand{\nou}[4]{\tensor[^{(#1)}]{#2}{^{#3}_{#4}}}

\newcommand{\I}{\mathrm{i}} %imaginaere Einheit
\newcommand{\E}{\mathrm{e}} %Euler Zahl
\newcommand{\CC}{\mathrm{cc.}} % komplex konjugiertes
\newcommand{\R}{\mathbb{R}}

\newcommand{\1}{\mathnormal{1}}

\newcommand{\eref}[1]{(\ref{#1})}

\DeclareMathAlphabet{\bbgreek}{U}{bbold}{m}{n}

\newcommand{\mtext}[1]{\text{\it #1}}
\usepackage{tikz-cd}

\newcommand\vpm{\mathbin{\vcenter{\hbox{
  \oalign{\hfil$\scriptstyle+$\hfil\cr
          \noalign{\kern-.3ex}
          $\scriptscriptstyle({-})$\cr}}}}}
\DeclareMathAlphabet{\sfit}{OT1}{fos}{sb}{it}
\DeclareMathAlphabet{\mathsf}{OT1}{fos}{sb}{n}

\definecolor{darkgreen}{rgb}{0.01, 0.75, 0.24}
\definecolor{darkblue}{rgb}{0.01, 0.24, 0.75}

\usepackage[multiple, flushmargin]{footmisc}

\let\originalleft\left
\let\originalright\right
\renewcommand{\left}{\mathopen{}\mathclose\bgroup\originalleft}
\renewcommand{\right}{\aftergroup\egroup\originalright}

\newcommand{\dbarvar}{{\mathrm{d}\mkern-7.5mu\lower.18ex\hbox{$\textasciitilde$}\mkern-1.5mu}}

\usepackage{ulem}

\renewcommand{\emph}[1]{{\it #1}}
\usepackage{mathtools}
\usepackage{empheq}
\usepackage[skins,theorems]{tcolorbox}
\tcbset{highlight math style={enhanced,
  colframe=red!60!black,colback=white,arc=4pt,boxrule=1pt}}
\definecolor{mygray}{RGB}{235, 235, 235}
\newcommand*\mygraybox[1]{%
\colorbox{mygray}{\hspace{1em}#1\hspace{1em}}}

\begin{document}

\begin{abstract}
\noindent An analysis is given of the local phase space of gravity coupled to matter to second order in perturbation theory. Working in local regions with boundaries at finite distance, we identify matter, Coulomb, and additional boundary modes. The boundary modes take the role of reference frames for both diffeomorphisms and internal Lorentz rotations. Passing to the quantum level, we identify the constraints that link the bulk and boundary modes. The constraints take the form of a multi-fingered Schrödinger equation, which determines the relational evolution of the quantum states in the bulk with respect to the quantum reference fields for the local gravitational symmetries at the boundary. Taking the boundary to infinity, we obtain quantum reference frames for asymptotic symmetries.

\end{abstract}%
\title{Quantum Reference Frames at the Boundary of Spacetime}
\author{Viktoria Kabel,${}^{1,2}$ \v{C}aslav Brukner,${}^{1,2}$  and Wolfgang Wieland${}^{1,2,3}$}
\address{${}^{1}$Institute for Quantum Optics and Quantum Information (IQOQI)\\Austrian Academy of Sciences\\Boltzmanngasse 3, 1090 Vienna, Austria
}
\address{${}^{2}$Vienna Center for Quantum Science and Technology (VCQ)\\Faculty of Physics, University of Vienna\\Boltzmanngasse 5, 1090 Vienna, Austria}

\address{${}^{3}$Institute for Quantum Gravity, Theoretical Physics III\\Friedrich-Alexander Universität Erlangen-Nürnberg, Department Physik\\Staudtstra\ss e 7/B2, 91058 Erlangen, Germany\\{\vspace{0.5em}\normalfont \today}}

\maketitle
\vspace{-1.2em}
\hypersetup{
  linkcolor=black,
  urlcolor=black,
  citecolor=black
}
{\tableofcontents}
%:
\hypersetup{
  linkcolor=black,
  urlcolor=darkblue,
  citecolor=darkblue,
}
\begin{center}{\noindent\rule{\linewidth}{0.4pt}}\end{center}\newpage

\section{Introduction}

\noindent To the best of our knowledge, all matter {obeys quantum mechanical laws}. The equivalence principle speaks of another such universality of nature. Gravity interacts with all matter in the same universal way. In the linearized regime with small perturbations around Minkowski space, the equivalence principle is a consequence of local gauge invariance and global momentum conservation \cite{Weinberg:1964ew}. 
 Besides the particles of the standard model, there is now the graviton propagating over a fixed background. 
 From this perspective, gravity behaves just like another form of matter. Thus, it should have a quantum description as well. The conceptually simplest approach towards such a quantum theory of gravity is to keep the background fixed and treat the perturbations as a quantum field on Minkowski space. There are two major problems with this viewpoint. The first problem is technical---the resulting theory is perturbatively nonrenormalizable, that is, it fails at high frequencies with respect to the fixed classical background. 
 The second problem is conceptual. The perturbative quantization leaves the background fixed, treating it as a classical arena on which the quantum fields evolve. 
 From the perspective of the full nonlinear Einstein equations, this seems highly problematic.
 The split $g_{ab}=\eta_{ab}+2f_{ab}$ of the curved spacetime metric $g_{ab}$ into a classical background $\eta_{ab}$, which has no quantum properties at all, and a perturbation $f_{ab}$ is completely arbitrary. If quantum theory is truly universal, it will apply to the gravitational field as a whole. The background would have quantum properties as well. While this possibility
 has been largely ignored in the perturbative approaches, it is the basic \emph{raison d'etre} of  nonperturbative approaches to quantum gravity.  \medskip

In this article, we present a novel {solution for} this second and more conceptual problem within the perturbative regime. {The key novelty is that we take into account the quantum properties of the background metric and encode them into a particular form of quantum reference frame at the boundary of spacetime.} In this way, our approach sits halfway between perturbative gravity and nonperturbative approaches. {Moreover, it connects two converging areas of research: the area of boundary modes}, which studies the state space of gauge theories in bounded regions of spacetime, e.g.\ \cite{DF16,Speranza18,CL21,CLP22,Freidel21,Freidel:2020xyx,FOPS21,Wieland21,Wieland:2021vef,Wieland:2017cmf,CarrozzaHoehn21,CEH22,GHK22}, and the field of quantum reference frames, which has regained much interest in recent years in the quantum foundations community, e.g.\ \cite{loveridge_symmetry_2018,Giacomini_2019,Giacomini_spin,vanrietvelde2018change,delaHamette2020,hoehn2019trinity,castroruiz2021relative,delahamette2021perspectiveneutral}.

The basic idea is motivated by the following observation. For any flat signature $(-$$+$$+$$+)$ metric $\eta_{ab}$ on a four-dimensional spacetime manifold $\mathcal{M}\simeq\R^4$ there always exist coordinate fields\footnote{That is, differentiable and invertible maps $X^\mu: \mathcal{M}\rightarrow\R^4$, whose inverse is also differentiable.} $\{X^\mu= (cT,X,Y,Z)\}$ that bring the metric into the standard diagonal Minkowskian form
\begin{equation}
\eta_{ab}=\eta_{\mu\nu}\partial_aX^\mu\partial_bX^\nu = -c^2 \partial_a T\partial_b T+\partial_a X\partial_b X+\partial_a Y\partial_b Y+\partial_a Z\partial_b Z.\label{flatmtrc}
\end{equation}

\noindent Since the metric is flat, we can refer to the four scalar fields $X^\mu$ as inertial frames of reference. More concretely, Eq.~\eqref{flatmtrc} can be read in two ways. If we know $\eta_{ab}$, we will also know $X^\mu$ modulo global symmetries (translations, boosts and rotations).  If, on the other hand, we take the fields $X^\mu$ as given, Eq.~\eref{flatmtrc} defines a flat reference metric $\eta_{ab}$ with respect to which the reference frame $\{X^\mu\}$ is inertial. {Thus, the metric $\eta_{ab}$ and the inertial reference fields $\{X^\mu\}$ contain the same information.} Combining the universality of quantum theory with the observation that all reference frames are, ultimately, instantiated by material objects, it is then natural to consider the possibility that the reference frames themselves have a quantum description. This is the idea underlying the research on \emph{quantum reference frames}. By the above argument, the resulting quantum coordinates define a flat Minkowskian quantum geometry $\widehat{\eta}_{ab}=\eta_{\mu\nu}\partial_a\widehat{X}^\mu\partial_b\widehat{X}^\nu $. Finding a quantum description of the background geometry and understanding  quantum inertial frames  are thus two sides of the same coin.

\begin{figure}[h!]
    \centering
    \includegraphics[scale=0.85]{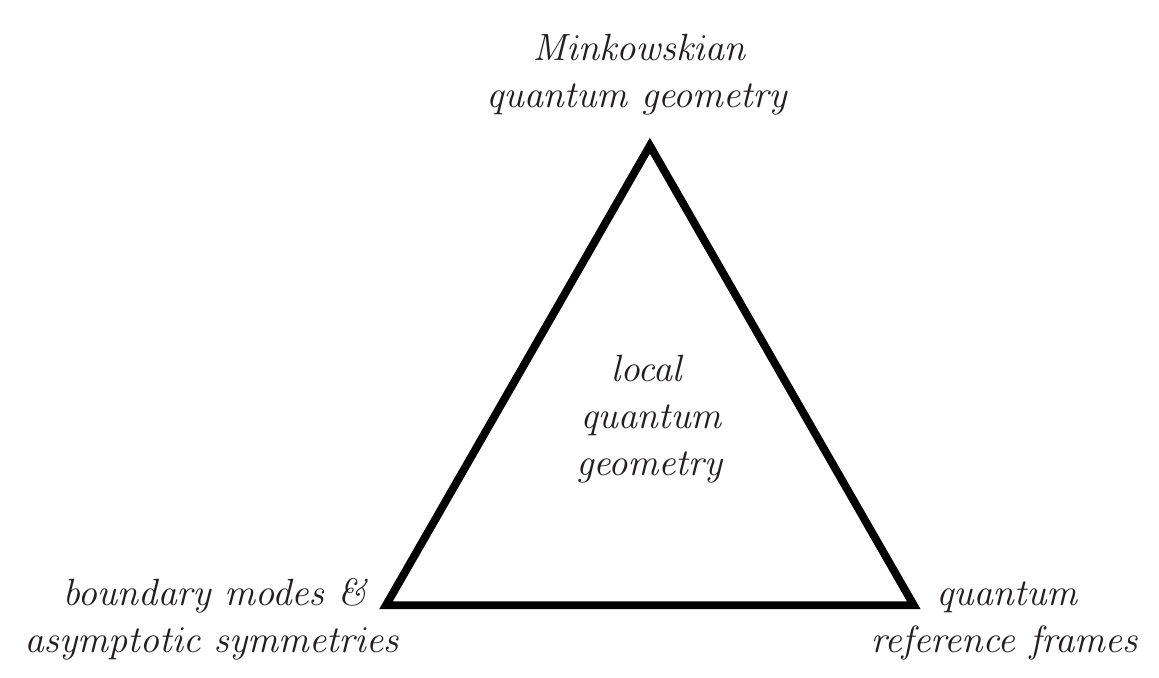}
    \caption{Basic themes of this paper. We consider linearized quantum gravity in finite regions (\emph{local quantum geometry}). At the boundary, we obtain additional edge modes that carry a representation of the boundary symmetry group (left corner). Going to an asymptotic region, we recover asymptotic Bondi--Metzner--Sachs symmetries. In the absence of radiation, the boundary modes take the form of inertial coordinate frames, which give rise to a notion of Minkowskian quantum background geometry (upper corner). Finally, the boundary modes serve as quantum reference frames for the relational evolution in the bulk (right corner).}
    \label{fig:triangle}
\end{figure}

From a general relativistic point of view, it seems odd at first to {place the focus}  on coordinates. 
{After all, the observables of the theory are coordinate invariant. 
General nonlinear coordinate transformations are mere gauge symmetries, mathematical redundancies that do not affect physical predictions. However, this is no longer true when we take into account the boundary conditions \cite{Carlip:1996yb}.} When we consider finite subsystems, or spacetimes with asymptotic or inner boundaries {such as a black hole horizon}, we need to impose boundary conditions {to obtain definite predictions}. Only those diffeomorphism that preserve the then-fixed boundary conditions remain gauge transformations. The residual diffeomorphisms that change the boundary conditions turn into actual symmetries, transforming one physical state into an inequivalent configuration. Gauge symmetries are thus broken at the boundary that separates the system from its environment. At the boundary, the coordinate fields  $X^\mu$ regain physical meaning and the value of physical observables will depend on them. As a result, we obtain additional boundary degrees of freedom. Since these boundary modes are fields intrinsic to a codimension-two surface bounding a partial Cauchy surface, they are often referred to as \emph{edge modes}. Classically, these edge modes can be interpreted as reference frames for the symmetry groups of the theory \cite{CarrozzaHoehn21,CEH22,GHK22}---an interpretation that extends to the quantum level{, where, as we will show, the edge modes manifest themselves as a type of quantum reference frame. The connection between the background metric, edge modes, and quantum reference frames, which is established in our work, is depicted in \hyperref[fig:triangle]{Fig.~1}.}\medskip 

To access all possible gravitational boundary modes, we  consider a compact region of space on an initial hypersurface. While there is a large and growing literature characterizing the phase space of such regions in general relativity and the emergence of additional reference degrees of freedom at the boundary \cite{DF16,Speranza18,HarlowWu20, CL21,CLP22,Freidel21,Freidel:2020xyx,FOPS21,Wieland21,Wieland:2021vef,Wieland:2017cmf,CarrozzaHoehn21,CEH22,GHK22} most of the analysis is performed within the full nonperturbative framework. For the present work, on the other hand, {we utilize perturbative methods}. This viewpoint has two main advantages. First of all, we obtain a clear physical interpretation of the bulk and boundary modes. Second, we can take the analysis to the quantum level without going into the mathematical difficulties of the nonperturbative approaches. Starting at the classical level, we consider the perturbative expansion of the presymplectic two-form with respect to the gravitational coupling constant, which is $\sqrt{8\pi G}$. To understand the effects of matter in a physically transparent model, we consider $N$ point particles contained within the region. 
Upon expanding the metric to second order in the perturbation, the presymplectic two-form splits into contributions from particles, gravitational waves, Coulomb fields sourced by both the particles and the gravitational waves themselves, as well as additional reference fields at the boundary. The reference frames take the form of coordinate fields and local $SO(1,3)$ transformations at the boundary, which provide a reference frame for internal boosts and rotations of the tetrads. The conjugate variables find their interpretation as boundary currents for energy, momentum and spin. 
At the quantum level, our analysis has three important implications. 
First of all, the partition of the phase space into matter, radiative, and boundary modes, together with the resulting Poisson commutation relations, informs us about the basic constituents of the algebra of observables at the quantum level. We will argue, in particular, that in any quantum theory of gravity the boundary modes must admit a quantum description as well, which leads immediately to the notion of \emph{quantum} reference frames. Second, as we will see below, the field equations place constraints on the relation between the bulk and boundary modes. The then-necessary constraints are a remnant of the Wheeler--DeWitt equation of the full nonperturbative theory and describe the relational evolution of the wavefunction with respect to the reference fields at the boundary. Finally, we take the limit to an asymptotic boundary and show how unitary representations between different asymptotic reference frames will carry a quantum representation of the group of asymptotic Bondi--Metzner--Sachs (BMS) transformations \cite{Sachs,Bondi21,AshtekarNullInfinity,Ashtekar:1978zz} at spatial infinity. Our results go beyond the existing framework for quantum reference frames, because they provide an extension to pointwise  transformations at the boundary and thereby establish the connection to asymptotic symmetries, which are infinite dimensional. At every point of the boundary, there sits a physical frame of reference for Poincaré translations, boosts and rotations. Since any full theory of quantum gravity should agree at least at low energies with the linearized theory, the insights gained from our analysis are expected to hold across different approaches.
\medskip

As for the remainder of this article, we will start by explaining in \hyperref[sec:edgemodes]{Sec.~\ref{sec:edgemodes}} the emergence of boundary modes in gauge theory and gravity using the example of electromagnetism. \hyperref[sec:ModesOfGravity]{Sec.~\ref{sec:ModesOfGravity}}  contains the main technical part of the paper. We review the perturbative expansion of the Einstein equations in the tetrad formulation before applying our analysis to linearized gravity in a bounded region. Readers that are not interested in the technical derivation may skip the first part of \hyperref[sec:ModesOfGravity]{Sec.~\ref{sec:ModesOfGravity}} and start with the summary at the end of \hyperref[subsec:PertbtvExpns]{Sec.~3.2}. After discussing the physical characterization of the bulk degrees of freedom---matter and radiation---we turn to the boundary modes in \hyperref[sec:boundaryModes]{Sec.~\ref{sec:boundaryModes}}.  
We close our analysis by summarizing and comparing our results to previous works and provide a short outlook in \hyperref[sec:conclusion]{Sec.~\ref{sec:conclusion}}

\section{Boundaries Break Gauge Symmetries}
\label{sec:edgemodes}

\noindent In the following, we choose a finite  domain $\mathcal{D}\subset\Sigma$ on a Cauchy surface $\Sigma$ and characterize its phase space. The boundary $\mathcal{C}=\partial\mathcal{D}$ separates the inside of the domain from the rest of the Cauchy surface. The presence of the boundary breaks the gauge symmetries. This is true both for the usual Yang--Mills type of gauge theories with internal symmetry groups, but it is also true for generally covariant theories, such as general relativity. In either case, new edge modes are excited at the boundary \cite{Balachandran:1994up,Carlip95,DF16}. In quantum theory, {edge modes provide a neat tool to characterize the quantum entanglement of physical states across a codimension-two surface. Therefore, such a boundary $\mathcal{C}=\partial\mathcal{D}$ is also often referred to as an \emph{entangling surface}.} An intuitive and often cited way to understand the emergence of such boundary modes is to consider Wilson loops in electromagnetism. This will be the focus of this section. In \hyperref[sec:ModesOfGravity]{Sec.~\ref{sec:ModesOfGravity}}, we turn to general relativity.\medskip

\begin{figure}[h!]
    \centering
    \includegraphics[scale=0.3]{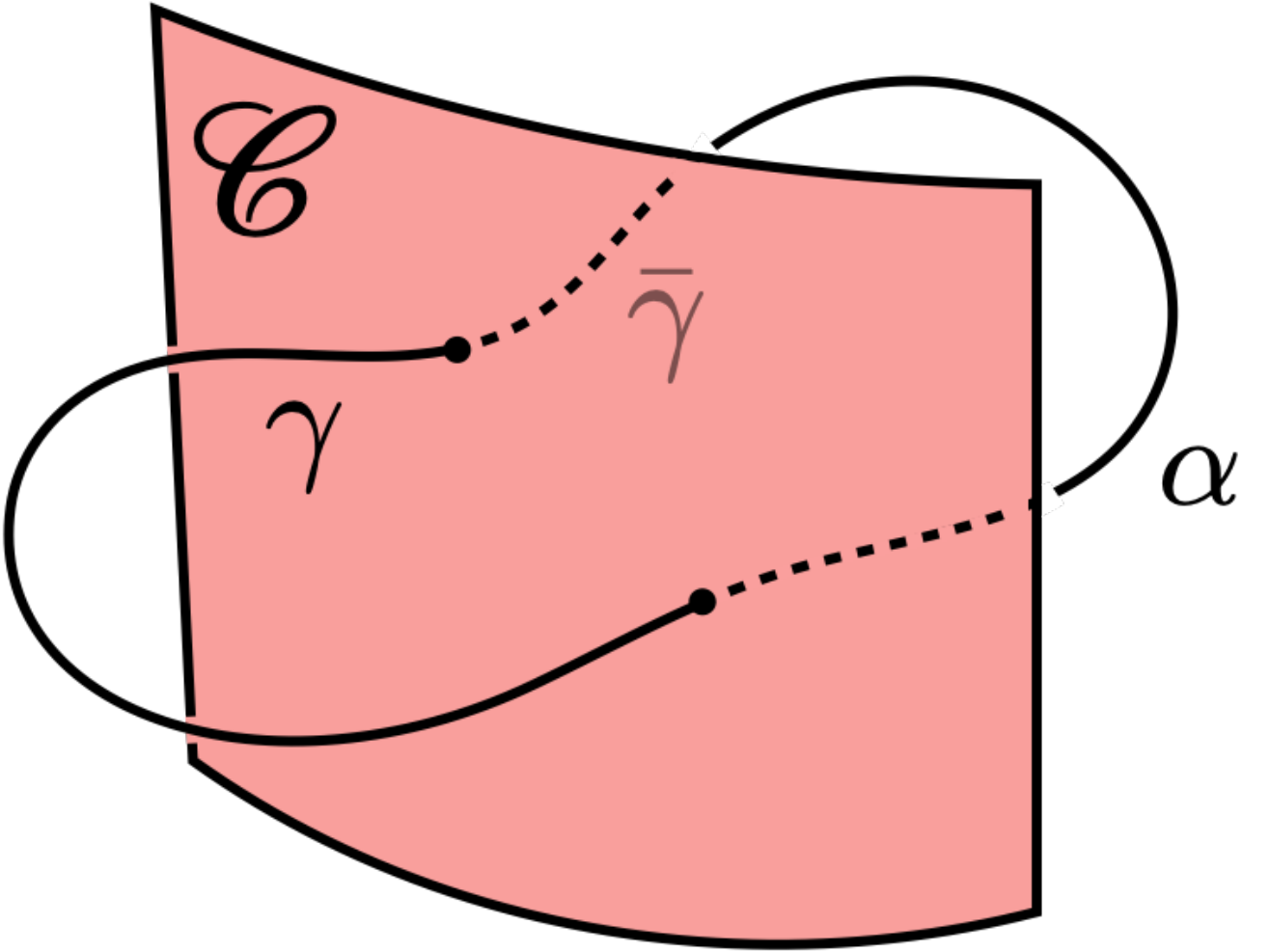}
    \caption{The entangling surface $\mathcal{C}$ cuts a Wilson line $\alpha$ into two parts $\gamma$ and $\bar{\gamma}$, breaking gauge invariance, {introducing} boundary modes at the puncture.}
    \label{fig:wilsonLoop}
\end{figure}

\noindent Consider thus a closed loop $\alpha:S^1\rightarrow\Sigma$ on the initial hypersurface. The Wilson loop observable
\begin{equation}
h_\alpha[A] = \E^{-\I\,q\,\oint_\alpha A }
\end{equation}
is invariant under $U(1)$ gauge transformations that shift $A$ into $A-\di\lambda$. This is no surprise because the gauge parameter $\lambda$ is not among the physical degrees of freedom in the bulk $\Sigma$. If, however, we restrict ourselves to a finite domain $\mathcal{D}$ and assume (see \hyperref[fig:wilsonLoop]{Fig.~2}) that the entangling surface $\mathcal{C}$ cuts the loop $\alpha$ into two separate paths $\gamma\subset\mathcal{D}$ and $\bar{\gamma}\subset\Sigma-\mathcal{D}$, the restriction of the Wilson loop observable to the compact domain $\mathcal{D}$ is no longer invariant under the  $U(1)$  transformations. Instead, we now find
\begin{equation}
h_\gamma[A-\di\lambda] = h_\gamma[A]\E^{\I\,q\, \sum_{p}\varepsilon_p\lambda(p)},
\end{equation}
where $\varepsilon_p$ is the relative orientation between the loop $\alpha$ and the entangling surface $\mathcal{C}$ at the puncture $p\in\alpha\cap\mathcal{C}$. In this sense, the gauge parameter $\lambda$ becomes physical at the boundary once we introduce an entangling surface that breaks the Wilson lines apart. The gauge parameter becomes a boundary mode dual to the charge excited at the puncture.\medskip

Another way to make the same observation is to consider the classical phase space of a bounded region in the \emph{covariant phase-space formalism} \cite{IyerWald94}. While we  point the reader to \cite{HarlowWu20,Ciambelli:2022vot,Freidel:2023bnj} for pedagogical introductions, we recall here the most important concepts and definitions that we need for the remainder of this work. If we start from the Lagrangian of the theory and introduce the phase space $\mathcal{P}$ via a Legendre transformation, the entire construction depends on a choice of time. This seems to be at odds with general covariance. The covariant phase-space formalism provides a way to reconcile the Hamiltonian description with covariance. Phase space $\mathcal{P}$ is built from the infinite-dimensional manifold of solutions to the field equations equipped with a \emph{symplectic form} $\Omega$, which is a two-form on the infinite-dimensional space. The two-form $\Omega$ provides a mapping between vectors and one-forms on field space (i.e.,\ it can be used to \emph{lower indices}). It does not, however, define a positive-definite inner product because it is skew-symmetric and may have degenerate null (i.e.,\ gauge) directions. 
Given a Lagrangian $\mathcal{L}$, the symplectic form can be derived from the first variation of $L$. Consider a theory of dynamical fields $\Phi^I$ on a $d$-dimensional manifold. The variation of the Lagrangian is given by
\begin{equation}
	\delta \mathcal{L} = E_I\delta \Phi^I + \mathrm{d}(\vartheta(\delta)),
\end{equation}
where $E_I=0$ are the Euler--Lagrange equations of motion and  $\vartheta$ denotes the symplectic potential. Notice that $\vartheta$ is a one-form on field space and a $(d-1)$-form on spacetime. It will be convenient to distinguish vector fields $\delta$ from exterior derivatives $\bbvar{d}$ on field space. When they act on the fundamental fields, the two are related via $\bbvar{d}\Phi^I(\delta) = \delta \Phi^I$. From the symplectic potential, one then constructs the \emph{symplectic current} by taking the exterior derivative $\bbvar{d}\vartheta$. The \emph{symplectic two-form} $\Omega$ derives from the latter through $\Omega = \int_\Sigma \bbvar{d} \vartheta$, where the integral is defined over a Cauchy hypersurface $\Sigma \subset \mathcal{M}$. Note that although the integral is defined over a Cauchy hypersurface, it is independent of which Cauchy hypersurface we choose, because $\mathrm{d}\bbvar{d} \vartheta = \bbvar{d}^2\mathcal{L}=0$ on shell. Thus, the choice of $\Sigma$ does not break the manifest covariance of the formalism. Finally, let us point out one important caveat in the derivation of the symplectic form given the Lagrangian: in the presence of symmetries, $\Omega$ is not invertible. It is therefore more correctly referred to as a \emph{pre-}symplectic form. To obtain an invertible symplectic form, one has to quotient out the symmetries of the phase space \cite{Marsden1974}. We will, however, simply work with the presymplectic form for the remainder of our discussion.

Returning to the example above, we are interested in the symplectic structure of a bounded region in electrodynamics. Considering the variation of the Lagrangian for Maxwell theory, $\mathcal{L}_{\mathrm{ED}}[A] = -\frac{1}{2}(F[A]\wedge \star F[A])$, where $F=\di A$ is the field strength and $\star$ is the Hodge dual, one finds that
\begin{equation}
    \delta \mathcal{L}_{\mathrm{ED}}[A] = \underset{E\delta A}{\underbrace{\delta A \wedge (\mathrm{d}\star F)}} + \underset{\mathrm{d}\vartheta(\delta)}{\underbrace{\mathrm{d}\left((\star F)\wedge \delta A\right)}}.
\end{equation}
To describe the phase-space structure of a \emph{bounded} region, we further restrict the integral of the presymplectic current to a compact domain $\mathcal{D}$. The resulting presymplectic two-form is
\begin{equation}
\Omega_{\mathcal{D}}  = \int_{\mathcal{D}}  \bbvar{d}[{\star}F]\wedge\bbvar{d} A,\label{EMOm}
\end{equation}
Consider now an arbitrary gauge-fixing condition such that we can write the vector potential $A$ as the sum of a gauge-fixed solution $A_{\mtext{gf}}$, which captures all the physical degrees of freedom in the bulk, and some arbitrary gauge transformation $\lambda$. In other words,
\begin{equation}
A = A_{\mtext{gf}}-\di\lambda.
\end{equation}
Inserting this decomposition back into \eref{EMOm}, and taking into account the field equations $\di\star F=0$, we obtain
\begin{equation}
\Omega_{\mathcal{D}}  = \int_{\mathcal{D}}  \bbvar{d}[{\star}F]\wedge\bbvar{d} A_{\mtext{gf}}-\oint_{\mathcal{C}} \bbvar{d}[{\star}F]\,\bbvar{d}\lambda.
\label{EMphasespace}
\end{equation}
The first term describes the {geometry of the gauge-invariant phase space for the electromagnetic field} in the region $\mathcal{D}$. {The boundary term describes the geometry of the edge modes. Its simple functional form tells us that the gauge parameter $\lambda$ is conjugate, in the phase-space sense,} to the electric field through the boundary. 

The goal of the next two sections is to provide an in-depth analysis of the gravitational case for linearized fields, where the analog of the boundary modes $\lambda$ will provide us with a local Poincar\'e frame at the boundary.\medskip

There is, however, a fundamental difference between the gravitational and the electromagnetic case. In electromagnetism, we can consider situations where there are no charged particles and the radiation field is confined to a finite pulse of compact support, localized within $\mathcal{D}$. For any such configuration, the boundary term in Eq.~\eref{EMphasespace} clearly vanishes. This is so because photons are charge neutral. They do not act as a source for radiation themselves. The gravitational case is fundamentally different. Gravitational radiation carries gravitational charge (energy, momentum, spin) and thus sources gravity as well. As soon as there is gravitational radiation in $\mathcal{D}$, it will also excite a surface charge at the boundary $\mathcal{C}$. As we will see below, this surface charge survives as a finite contribution to the presymplectic structure at the perturbative level.

\section{Modes of Gravity: Radiative Fields, Coulombic Fields, Boundary Modes}
\label{sec:ModesOfGravity}

\subsection{Review: Perturbative expansion of the field equations in the tetrad formalism}
\noindent Let us now turn to the general relativistic case. Our starting point is gravity in the first-order tetrad formalism. The analysis could be done equally well in the metric formalism. One important difference is that working with tetrads, we obtain an additional gauge redundancy in the bulk, namely internal $SO(1,3)$ frame rotations of the frame field. This leads to additional reference fields for the Lorentz group at the boundary. These would be absent in metric gravity. In the context of this paper, the main advantage of working with tetrads rather than the metric is that tetrads are tailor-made to deal with how boundaries couple to the fields inside the manifold.\footnote{Tetrads are differential forms. Given a differential form in the bulk, there is a natural projection onto the boundary, namely the pullback. The pullback requires only an embedding, not a metric. Hence, it provides a more primitive way to introduce fields intrinsic to the boundary no matter whether the boundary is spacelike, timelike, or null.}  In tetrad gravity, the geometry is encoded into an orthonormal frame $e^\alpha$ that determines the metric tensor:
\begin{equation}
g_{ab} = \eta_{\alpha\beta}\ou{e}{\alpha}{a}\ou{e}{\beta}{b}.
\end{equation}
In addition, there is the spin connection $\ou{A}{\alpha}{\beta}$ with corresponding covariant exterior derivative $\nabla = \di +[A,\cdot]$. Here, internal Lorentz indices $\alpha,\beta,\dots$ are raised and lowered with the flat $(-$$+$$+$$+)$ Minkowski metric $\eta_{\alpha\beta}$, $a,b,c,\dots$ are abstract spacetime indices and $[\cdot,\cdot]$ is the Lie bracket for the Lie algebra $\mathfrak{so}(1,3)$. In this language, solving the Einstein equations amounts to imposing the equations
\begin{align}
{\ast}\ou{F}{\alpha}{\beta}\wedge e^\beta & = 8\pi G\, T^\alpha,\label{EEq1}\\
\nabla e^\alpha  = 0,\label{EEq2}
\end{align}
where $T^\alpha$ is the energy-momentum three-form of matter, which is the Hodge dual of the usual stress energy tensor $T_{ab}$, i.e.,\
\begin{equation}
\ou{T}{\alpha}{abc} = -\ou{e}{\alpha}{d}T^{df}\varepsilon_{fabc}.
\end{equation}
In addition, $\ast \ou{F}{\alpha}{\beta}$ denotes the \emph{internal} Hodge dual of the $\mathfrak{so}(1,3)$-valued curvature two-form, i.e.,\ $\ast F_{\alpha\beta}=\frac{1}{2}\epsilon_{\alpha\beta\gamma\delta} F^{\gamma\delta}$, with $\ou{F}{\alpha}{\beta}=\di\ou{A}{\alpha}{\beta}+\ou{A}{\alpha}{\gamma}\wedge\ou{A}{\gamma}{\beta}$ denoting the curvature of the connection {and $\epsilon_{\alpha\beta\gamma\delta}$ is the Levi-Civita tensor with respect to internal indices with conventions $\epsilon_{0123}=1$. Furthermore, $\varepsilon_{abcd}=\epsilon_{\alpha\beta\gamma\delta}\ou{e}{\alpha}{a}\ou{e}{\beta}{b}\ou{e}{\gamma}{c}\ou{e}{\delta}{d}$ is the usual Levi-Civita tensor in spacetime indices.}\medskip

In flat space, with vanishing energy momentum tensor, i.e.,\ $T_{ab}=0$, the general solution to the field equations is simply given by
\begin{align}
e^\alpha &= \ou{\Lambda}{\alpha}{\mu}\di X^\mu,\label{Minkowski1}\\
 \ou{A}{\alpha}{\beta} & = \ou{\Lambda}{\alpha}{\mu}\di\uo{\Lambda}{\beta}{\mu},\label{Minkowski2}
\end{align}
where the coordinate functions  $X^\mu:\mathcal{M}\rightarrow\R^4$ determine an inertial frame of reference on spacetime $\mathcal{M}$ and $\ou{\Lambda}{\alpha}{\mu}:\mathcal{M}\rightarrow SO(1,3)$ is a local Lorentz transformation.
In the following, we consider a perturbative expansion of the field equations around such Minkowskian solutions. We are interested in a decoupling limit with weak interaction and self-interaction among gravity and matter.

The perturbative expansion is written as a formal power series with respect to the coupling constant $\ell=\sqrt{8\pi G}$, which is the Planck length. We thus introduce a perturbation
\begin{align}
e^\alpha &= \ou{\Lambda}{\alpha}{\mu}\big(\di X^\mu + f^\mu\big),\label{eexpans}\\
 \ou{A}{\alpha}{\beta} & = \ou{\Lambda}{\alpha}{\mu}\di\uo{\Lambda}{\beta}{\mu}+\ou{\Lambda}{\alpha}{\mu}\ou{\Delta}{\mu}{\nu}\uo{\Lambda}{\beta}{\mu}\label{Aexpans}.
\end{align}
and expand the remaining dynamical fields, i.e.,\ the tetrad perturbation $f^\mu$, the difference tensor $\ou{\Delta}{\mu}{\nu}$, and the energy momentum three-form $T^\mu$ into a formal power series with respect to the coupling constant $\ell$. We thus write
\begin{align}
 f^\mu &=\sum_{n=1}^\infty \ell^n\,\nT{n}{f}{^\mu},\label{fseries}\\
 \Delta^{\mu}{}_{\nu} & = \sum_{n=1}^\infty \ell^n\,\tensor[^{(n)}]{\Delta}{^\mu_\nu},\label{Dseries}\\
 T^\mu & = \sum_{n=0}^\infty \ell^{n} \,\tensor[^{(n+2)}]{T}{^\mu}.\label{Tseries}
\end{align}
Notice that the expansion of the energy-momentum three-form starts out one order lower than the expansion for $f^\mu$ and $\ou{\Delta}{\mu}{\nu}$. This is so because at lowest order in the expansion, the energy-momentum tensor is independent of the metric perturbation.

It is then also useful to decompose the tensorial coefficients that appear in this expansion, \eref{fseries} and \eref{Dseries}, in terms of the zeroth order of the frame field, i.e.,\ we write
\begin{align}
\nT{n}{f}{^\mu} &= \nT{n}{f}{^\mu_\nu}(X)\,\di X^\nu,\\
\nT{n}{\Delta}{^\mu_\nu}&=\nou{n}{\Delta}{\mu}{\nu\rho}(X)\,\di X^\rho.
\end{align}

Here, all indices $\mu,\nu,\rho,\dots$ are treated as internal, i.e.,\ they are raised and lowered with the signature $(-$$+$$+$$+)$ flat  Minkowski metric $\eta_{\mu\nu}$, e.g.\ $\nT{n}{f}{_{\mu\nu}}=\eta_{\mu\rho}\nou{n}{f}{\rho}{\nu}$. Note that, at each order in the perturbation, any antisymmetric part of $\nT{n}{f}{_{\mu\nu}}$ can be reabsorbed back into a redefinition of the local Lorentz transformation $\ou{\Lambda}{\alpha}{\mu}$. Hence, we  can always assume without loss of generality
\begin{equation}
\nT{n}{f}{_{[\mu\nu]}}=0.
\end{equation}

We are now ready to consider the perturbative expansion of the field equations. At lowest order $n=1$, the field equations are
\begin{align}
{\ast}\ou{{\big[\di\nT{1}{\Delta}{}\big]}}{\mu}{\nu}\wedge \di X^\nu & = 0\Leftrightarrow \partial_\rho\nT{1}{\Delta}{^{\rho\nu}_{\mu}}-\partial_\mu\nT{1}{\Delta}{^{\rho\nu}_\rho}=0,\label{o1EOM1}\\
\di\nT{1}{f}{^\mu}+\nou{1}{\Delta}{\mu}{\nu}\wedge\di X^\nu &=0\Leftrightarrow \nT{1}{\Delta}{_{\mu\nu}}=-2\partial_{[\mu}\nT{1}{f}{_{\nu]}}\label{o1EOM2},
\end{align}
where the operator $\ast$ only acts on internal indices, i.e.,\ $\ast\ou{{\big[\di\nT{1}{\Delta}{}\big]}}{\mu}{\nu}=\frac{1}{2}\ou{\varepsilon}{\mu}{\nu\rho\sigma}\big(\partial_\lambda\nou{1}{\Delta}{\rho\sigma}{\tau}\big)\,\di X^\lambda\wedge\di X^\tau$. At next to leading order $n=2$, the field equations are
\begin{align}
{\ast}\ou{{\big[\di\nT{2}{\Delta}{}\big]}}{\mu}{\nu}\wedge \di X^\nu & = \nT{2}{T}{^\mu}+\nT{2}{t}{^\mu},\label{o2EOM1} \\
\di\nT{2}{f}{^\mu}+\nou{2}{\Delta}{\mu}{\nu}\wedge\di X^\nu &=\nT{2}{\theta}{^\mu},\label{o2EOM2} 
\end{align}
where we introduced the effective energy-momentum three-form and torsion two-form of the order $n=1$ perturbation, i.e.,\
\begin{align}
\nT{2}{t}{^\mu} & = -{\ast}\ou{{\big[\di\nT{1}{\Delta}{}\big]}}{\mu}{\nu}\wedge \nT{1}{f}{^\nu}-\frac{1}{2}{\ast}\ou{{}{\big[\nT{1}{\Delta}{},\nT{1}{\Delta}{}\big]}}{\mu}{\nu}\wedge\di X^\nu,\label{Tdef1}\\
\nT{2}{\theta}{^\mu}&=-\nou{1}{\Delta}{\mu}{\nu}\wedge\nT{1}{f}{^\nu}.\label{thetadef}
\end{align}
Here, $[\cdot,\cdot]$ denotes the Lie bracket between $\mathfrak{so}(1,3)$-valued $p$-forms, e.g.\ $\ou{[\Delta,\Delta]}{\mu}{\nu}=2\ou{\Delta}{\mu}{\rho}\wedge\ou{\Delta}{\rho}{\nu}$. The source terms on the right-hand side of \eref{o2EOM1} and \eref{o2EOM2} are a consequence of  the self-interaction of gravity---perturbations of the gravitational field at first order are themselves a source of gravity at second order.\medskip

The linearized field equations of the coupled system can now be solved in the usual way, see, e.g.,\ \cite{Poisson:2004ti} for a detailed analysis with pointlike sources.  Taking advantage of the gauge freedom of the field equations,\footnote{A gauge transformation of $f_{\mu\nu}$ can be always reabsorbed (order by order in the coupling constant) into a shift of the coordinate scalars $X^\mu$ and a redefinition of the Lorentz gauge transformation $\ou{\Lambda}{\alpha}{\mu}$. The perturbation \eref{modexpns} satisfies the usual transverse traceless gauge conditions $\partial_\mu\nT{1}{f}{^\mu_\nu}=0$ and $\nT{1}{f}{^\mu_\mu}=0$. } the general solution of  
(\ref{o1EOM1}, \ref{o1EOM2}) is a superposition of plane waves with amplitudes $a_\pm(\vec{k})$, polarization tensors $m_\mu m_\nu$, and four-momentum $k^\mu$,
\begin{equation}
\nT{1}{f}{_{\mu\nu}} = \frac{1}{{(2\pi)}^{\frac{3}{2}}}\int \frac{d^3 k}{2|\vec{k}|} \Big(m_\mu m_\nu\,a_+(\vec{k})\E^{\I k_\mu X^\mu}+\bar{m}_\mu \bar{m}_\nu\,a_-(\vec{k})\E^{\I k_\mu X^\mu}+\CC\Big),\label{modexpns}
\end{equation}
where, in terms of, e.g.,\ stereographic coordinates  $(z,\bar{z})$ on the two-sphere in momentum space,
\begin{equation}
k^\mu=\begin{pmatrix}
|\vec{k}|\\\vec{k}
\end{pmatrix}=
\frac{|\vec{k}|}{1+|z|^2}\begin{pmatrix}
1+|z|^2\\z+\bar{z}\\-\I(z-\bar{z})\\|z|^2-1
\end{pmatrix}
,\quad
m^\mu=\frac{1}{\sqrt{2}}\frac{1}{1+|z|^2}\begin{pmatrix}0\\1-z^2\\\I(1+z^2)\\2z\end{pmatrix}.
\end{equation}

To solve the second-order equations \eref{o2EOM1} and \eref{o2EOM2} by the method of Green's functions, it is useful to split the perturbation of the spin connection into two terms that take values in $\mathfrak{so}(1,3)$,
\begin{equation}
\nou{2}{\Delta}{\mu}{\nu} = \nou{2}{\Gamma}{\mu}{\nu} + \nou{2}{C}{\mu}{\nu}.
\end{equation}
Here, the first terms is defined as the solution of the equation
\begin{equation}
\di\nT{2}{f}{^\mu} + \nou{2}{\Gamma}{\mu}{\nu}\wedge \di X^\nu = 0,
\end{equation}
which can be solved algebraically by
\begin{equation}
\nT{2}{\Gamma}{^\mu_\nu}=-\partial^{\mu}\nT{2}{f}{_\nu}+\partial_{\nu}\nT{2}{f}{^\mu}.
\end{equation}
In the same way, the components of $\nou{2}{C}{\mu}{\nu}=\nou{2}{C}{\mu}{\nu\rho}\di X^\rho$ can be obtained algebraically by inverting \eref{thetadef}. Writing $\nT{2}{\theta}{^\mu}=\frac{1}{2}\nT{2}{\theta}{^\mu_{\nu\rho}}\di X^\nu\wedge\di X^\rho$ for component functions $\nT{2}{\theta}{^\mu_{\nu\rho}}=-\nT{2}{\theta}{^\mu_{\rho\nu}}$, we obtain
\begin{align}
\nT{2}{C}{_{\rho\mu\nu}} =  -\nT{2}{\theta}{_{\rho\mu\nu}}+\nT{2}{\theta}{_{\mu\rho\nu}}+\nT{2}{\theta}{_{\nu\rho\mu}}.
\end{align}
We thus have
\begin{equation}
{\ast}\ou{{\big[\di\nT{2}{\Gamma}{}\big]}}{\mu}{\nu}\wedge \di X^\nu = \nT{2}{T}{^\mu}+\nT{2}{t}{^\mu} - {\ast}\ou{{\big[\di\nT{2}{C}{}\big]}}{\mu}{\nu}\wedge \di X^\nu  \equiv \tensor*[^{(2)}]{T}{^\mu_{\mtext{eff}}}.\label{secondEEq}
\end{equation}
Given the standard Lorentz gauge conditions
\begin{equation}
\partial_\rho\nT{2}{f}{^\rho_\mu}-\frac{1}{2}\partial_\nu\nT{2}{f}{^\rho_\rho}=0,\label{gaugecond}
\end{equation}
Eq.~\eref{secondEEq} amounts to the Einstein equations at second order in the perturbation, i.e.,\
\begin{align}\nonumber
\square \nT{2}{f}{_{\mu\nu}}&=
-\Big(\tensor*[^{(2)}]{T}{_{\mu\nu}^{\mtext{eff}}}-\frac{1}{2}\eta_{\mu\nu}\tensor*[^{(2)}]{T}{_{\rho}^{\mtext{eff}\,\rho}}\Big),\\
\tensor*[^{(2)}]{T}{_{\mu\nu}^{\mtext{eff}}}&=\frac{1}{3!}\tensor*[^{(2)}]{T}{_{\mu\rho\sigma\tau}^{\mtext{eff}}}\uo{\varepsilon}{\nu}{\rho\sigma\tau}.
\end{align}
where $\tensor*[^{(2)}]{T}{_{\mu\rho\sigma\tau}^{\mtext{eff}}}=\tensor*[^{(2)}]{T}{_{\mu[\rho\sigma\tau]}^{\mtext{eff}}}$ are the components of the effective order $n=2$ energy-momentum three-form and $\square=\partial_\mu\partial^\mu$ is the D'Alembertian. Assuming no incoming radiation from past null infinity at order $n=2$, the solution to \eref{secondEEq} for the gauge conditions \eref{gaugecond} is thus given by
\begin{equation}
\nT{2}{f}{_{\mu\nu}}(X) = \frac{1}{4\pi} \int d^4X' G_R(X-X') \Big(\tensor*[^{(2)}]{T}{_{\mu\nu}^{\mtext{eff}}}(X')-\frac{1}{2}\eta_{\mu\nu}\tensor*[^{(2)}]{T}{_{\rho}^{\mtext{eff}\,\rho}}(X')\Big),\label{2fsourcedby}
\end{equation}
where $G_R(X)=\delta(X^0-|\vec{X}|)/|\vec{X}|$ is the retarded Green function. 

\subsection{Perturbative expansion of the symplectic structure in a finite region}\label{subsec:PertbtvExpns}

\noindent We now turn to the perturbative expansion of the presymplectic two-form. The gravitational sector is governed by the Hilbert-Palatini action:
\begin{equation}
S_{\mtext{grav}}[A,e] = \frac{1}{16\pi G} \int_{\mathcal{M}}\ast(e_\alpha\wedge e_\beta)\wedge F^{\alpha\beta}[A].
\end{equation}
To add explicit matter sources, we add the action for $N$ point particles minimally coupled to the gravitational field. The resulting field equations are at best formal. The reason is intuitively clear \cite{Gralla:2008fg}. If we try to concentrate a given mass into a point, we will eventually create a black hole before reaching the point particle limit. A possibility to circumvent this problem is to shrink down the body and take at the same time the ultrarelativistic limit, where the mass goes to zero, but the energy, with respect to some preferred observer at infinity, is kept fixed. In this way one \emph{can} find exact distributional solutions to Einstein's equations; see for instance  \cite{Aichelburg1971,Balasin:1993kf}. In the following, we are dealing with perturbative gravity, where we can safely ignore this problem and work with  distributional point sources. We can then formally solve the  Einstein equations order by order in the coupling constant. 

The action for the coupled system  is the sum of the gravitational action and the action for the $N$ point particles. The total action is thus given by
\begin{equation}
S\big[A,e,\{\gamma_i,N_i,p_i\}_{i=1}^N\big] = S_{\mtext{grav}}[A,e] +\sum_{i=1}^N\int_{\gamma_i}\Big(p_\alpha^i e^\alpha-\frac{N_i}{2}\big(p^i_\alpha p^\alpha_i - m^2_i\big)\Big),\label{totalactn}
\end{equation}
where $p_\alpha^i$ is the four-momentum of the $i$th particle, $\gamma_i$ denotes its worldline in $\mathcal{M}$, and $N_i$ denotes a Lagrange multiplier that  imposes the mass shell condition with rest mass $m_i$. The resulting energy momentum three-form for the matter Lagrangian is distributional and given by the variation of \eref{totalactn} with respect to the tetrad. That is,
\begin{equation}
\ou{T}{\alpha}{abc} = \sum_{i=1}^N\int_{\gamma_i}\di \tau\,p^\alpha_i(\tau)\,\tilde{\delta}^{(4)}_{\gamma_i(\tau)}\utilde{\varepsilon}_{abcd}\partial^d_\tau,\label{Tdef2}
\end{equation}
where $\tilde{\delta}^{(4)}_p$ is the four-dimensional Dirac delta distribution peaked at $p\in\mathcal{M}$ (a scalar density of weight one) and $\utilde{\varepsilon}_{abcd}$ is the inverse Levi-Civita tensor density.

Next, we consider a perturbative expansion around the flat Minkowskian solution given in Eqs.~\eref{Minkowski1} and \eref{Minkowski2}. We write 
$p^\alpha = \ou{\Lambda}{\alpha}{\mu}p^\mu$, letting indices $\alpha,\beta,\gamma,\dots$ refer to the tetrad $e^\alpha$, wile indices $\mu,\nu,\dots$ from the second half of the Greek alphabet refer to the flat background coordinates $\{X^\mu\}$. We expand the particle action in powers of the coupling constant $\ell=\sqrt{8\pi G}$ and obtain
\begin{equation}
\int_{\gamma}\Big(p_\alpha  e^\alpha-\frac{N}{2}\big(p_\alpha p^\alpha - m^2\big)\Big) = \int_{\mathbb{R}}\Big(p_\mu\di q^\mu+\ell p_\mu\nT{1}{f}{^\mu_\nu}(q)\di q^\nu-\frac{N}{2}\big(p_\mu p^\mu +m^2\big)\Big)+\mathcal{O}(\ell^2),
\end{equation}
where $q^\mu(\tau)$ denotes the coordinates of the trajectory $\gamma:\tau\in\R\rightarrow\mathcal{M}\ni\gamma(\tau)$ with respect to the coordinate frame $\{X^\mu\}$. If we want to solve the field equations order by order in $\ell$, we would now also need to consider a perturbative expansion of the particle trajectory $q^\mu(\tau)$ in powers of the coupling constant, i.e.,\ $q^\mu(\tau)=\nT{0}{q}{^\mu}(\tau)+\ell\nT{1}{q}{^\mu}(\tau)+\dots$, see for example \cite{Gralla:2008fg}, where such an approach is developed in great detail to first order in $\ell$. In this paper, we are  only interested in the zeroth leading order of the symplectic structure. The first contribution to the symplectic structure from the particle trajectory is $\mathcal{O}(\ell^0)$. Hence, it is not required for the present purpose to consider higher-order perturbations of the  trajectories.\medskip

The first variation of the action gives the Einstein equations with a distributional energy momentum tensor \eref{Tdef2} in the bulk plus a boundary term. The boundary term determines the presymplectic current $\vartheta$. Schematically, we have
\begin{equation}
\delta[S] = \mathrm{EOM}(\delta) + \oint_{\partial \mathcal{M}}\vartheta(\delta),
\end{equation}
where the term $\mathrm{EOM}(\delta)$, which is linear in the field variation $\delta$, vanishes provided the equations of motion (\ref{EEq1}, \ref{EEq2}) are satisfied. 
Given the total action \eqref{totalactn} of the coupled system, the presymplectic current $\vartheta$ is given by\footnote{Notice that the first-order action \eref{totalactn} is linear in derivatives, i.e.,\ $S=\int\di t(P\dot{Q}-N^\mu C_\mu(P,Q))$ such that it is immediate to identify the symplectic potential $\vartheta=P\bbvar{d}Q$ and constraints $C_\mu(P,Q)\approx 0$. }
\begin{equation}
\vartheta_{abc} = \frac{3}{16\pi G}\,\epsilon_{\alpha\beta\gamma\delta}\,\ou{e}{\alpha}{[a}\ou{e}{\beta}{b}\bbvar{d}\ou{A}{\gamma\delta}{c]}+\sum_{i=1}^N \int_{\gamma_i}\di\tau\,\tilde{\delta}^{(4)}_{\gamma_i(\tau)}\,\partial^d_\tau\utilde{\varepsilon}_{dabc}\, p^i_f\bbvar{q}^f_i,\label{theta3form}
\end{equation}
where $ p^i_a= p^i_\alpha(\tau)\ou{e}{\alpha}{a}\big|_{\gamma_i(\tau)}$ is the four-momentum of the $i$th particle and
\begin{equation}
%:
\bbvar{q}_i^a = \left[\frac{\partial}{\partial X^\mu}\right]^a_{\gamma_i(\tau)}\bbvar{d}q^\mu_{i}(\tau)
\end{equation}
denotes the variation of its position, which defines a $T_{\gamma_i}\mathcal{M}$-valued one-form on configuration space.\medskip

\begin{figure}
\centering
    \subfigure[$(2+1)$-dimensional representation.]{
        \centering
        \includegraphics[scale=0.27]{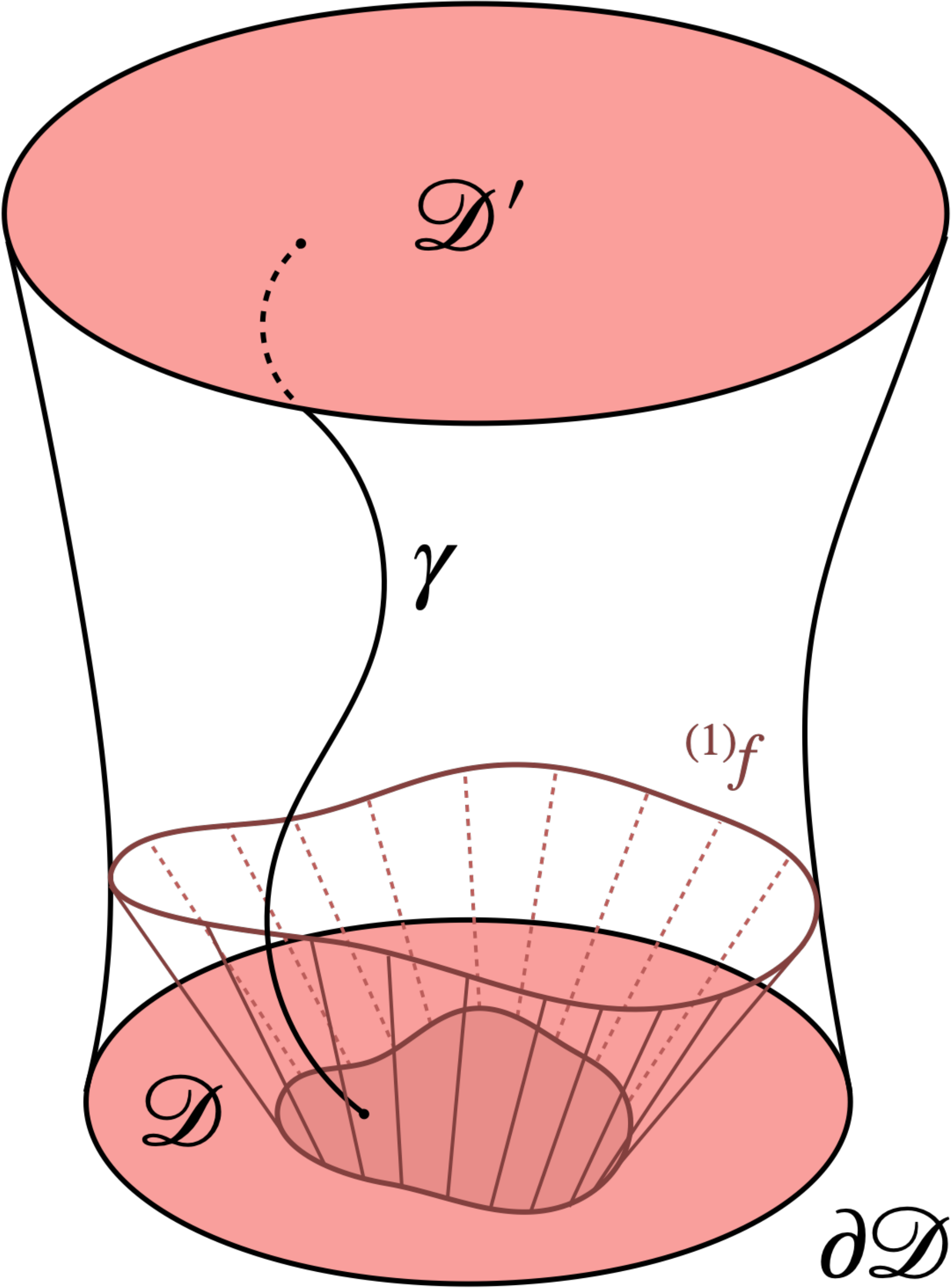}
        \label{fig:subregion3d}
    }
    \hspace{2cm}
    \subfigure[Penrose diagrammatic representation.]{
        \centering
        \includegraphics[scale=0.25]{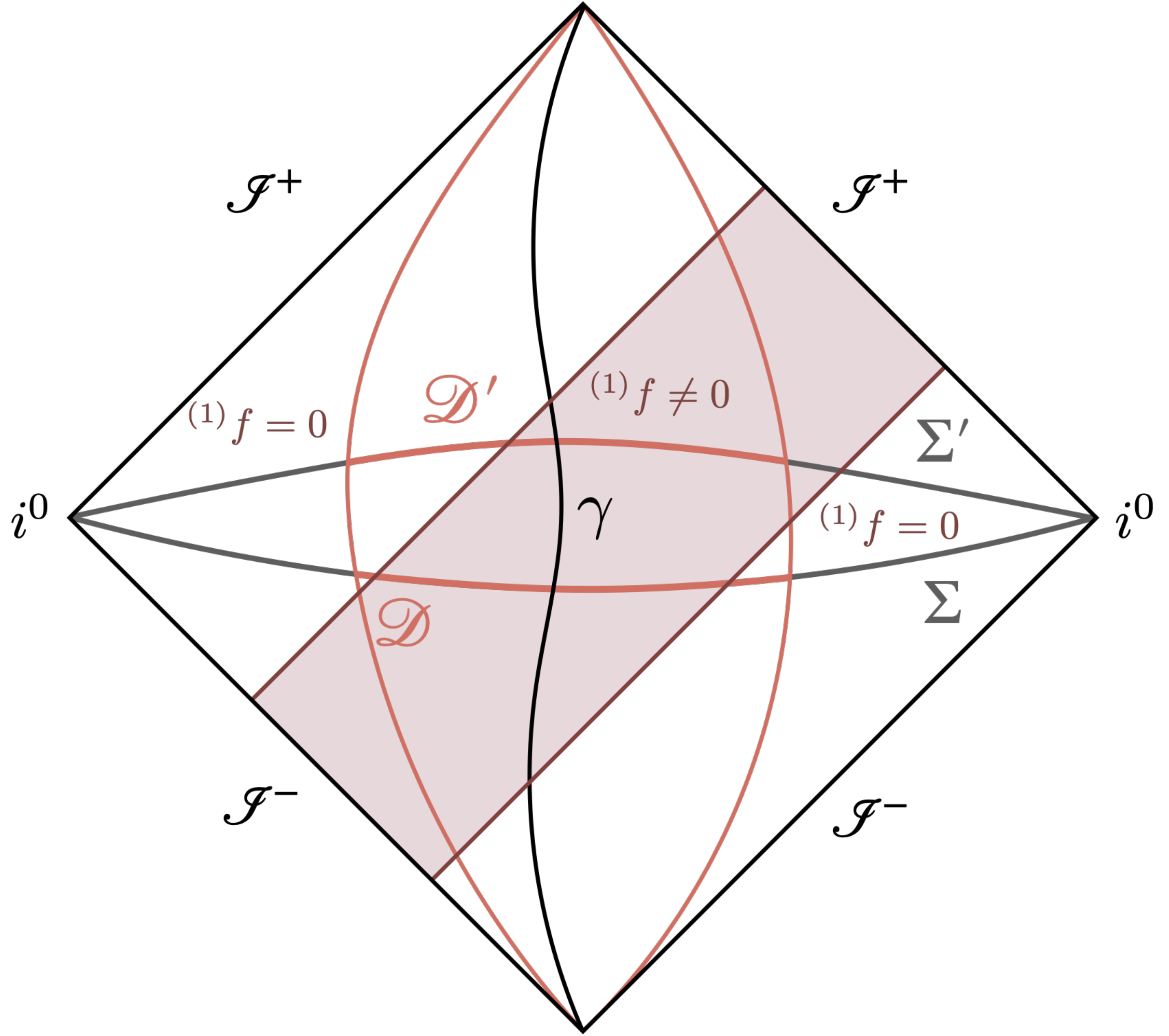}
        \label{fig:penrose}
    }
    \caption{A subregion of spacetime, delimited by a timelike cylinder and two subregions $\mathcal{D}$ and $\mathcal{D'}$ on Cauchy hypersurfaces $\Sigma$ and $\Sigma'$ respectively.  
    The worldline of a single point particle within this region is depicted by $\gamma$. In this section, we determine the bulk and boundary phase space on $\mathcal{D}$ in the linearized regime. In the derivation of the presymplectic form, the first-order perturbation of the tetrad field $^{(1)}f^\mu$, which accounts for gravitational waves, is assumed to have compact support within $\mathcal{D}$. An example for such a subregion of spacetime is represented by the shaded area in the Penrose diagram. In the white areas, $\nT{1}{f}{^\mu}=0$, but $\nT{2}{f}{^\mu}\neq0$.}
    \label{fig:subregion}
\end{figure}

Our goal in this section is to establish a full characterization of the gravitational phase space in a compact domain $\mathcal{D}$ in perturbative gravity. 
We choose the region $\mathcal{D}$ large enough such that it contains all particles $i=1,\dots,N$ and such that the $n=1$ radiative modes ${}^{(1)}f$ have compact support within $\mathcal{D}$. This simplifying assumption is illustrated in \hyperref[fig:subregion]{Fig.~\ref{fig:subregion}}.
Next, we integrate the presymplectic potential $\vartheta$ along  $\mathcal{D}$ and identify the contributions from matter and radiation and the additional boundary modes. Taking into account the torsionless condition \eref{EEq2} and the expansion (\ref{eexpans}, \ref{Aexpans}) of the configuration variables around the Minkowski solution, we can collect all variations of $\ou{\Lambda}{\alpha}{\mu}$
 into a boundary term
 \begin{align}\nonumber
\Theta_{\mathcal{D}} := \int_{\mathcal{D}}\vartheta = & \sum_{i=1}^N p_\mu^i\bbvar{d}q^\mu_i
+ \frac{1}{2\ell}\int_{\mathcal{D}}{\ast}\left(\di X_\mu\wedge\di X_\nu\right)\wedge\bbvar{d}\nT{1}{\Delta}{^{\mu\nu}}
+\frac{1}{2}\int_{\mathcal{D}}{\ast}\left(\di X_\mu\wedge\di X_\nu\right)\wedge\bbvar{d}\nT{2}{\Delta}{^{\mu\nu}}+\nonumber\\
&+\int_{\mathcal{D}}{\ast}\left(\di X_{[\mu}\wedge\nT{1}{f}{_{\nu]}}\right)\wedge\bbvar{d}\nT{1}{\Delta}{^{\mu\nu}}
+\frac{1}{2}\oint_{\partial\mathcal{D}} S_{\mu\nu}(\bbvar{d}\uo{\Lambda}{\rho}{\mu})\Lambda^{\rho\nu}+\mathcal{O}(\ell^2),\label{ThetaD1}
\end{align}
where $S_{\mu\nu}$ is the $\mathfrak{so}(1,3)$ spin current at the boundary:
\begin{equation}
S_{\mu\nu} := \frac{1}{\ell^2}\varphi^\ast_{\partial\mathcal{D}}\Big[{\ast}\left(\di X_{[\mu}\wedge\di X_{\nu]}\right)+2\ell^2{\ast}\left(\di X_{[\mu}\wedge\nT{2}{f}{_{\nu]}}\right)\Big].\label{Sdef}
\end{equation}
The fact that the $SO(1,3)$ gauge element $\ou{\Lambda}{\alpha}{\mu}$ appears only through a boundary term is a consequence of the $SO(1,3)$ gauge invariance of the action.\medskip

Let us now simplify the expression \eref{ThetaD1} further. Our initial assumption was that the free radiative data $\nT{1}{f}{^\mu}$ has compact support within $\mathcal{D}$. All boundary terms containing $\nT{1}{f}{^\mu}$ will thus vanish. It is for this reason, in fact, that no $\mathcal{O}(\ell^{-1})$ term appears in the definition of the $SO(1,3)$ spin current $S_{\mu\nu}$; see \eref{Sdef}. For the same reason, the second term in \eref{ThetaD1} also vanishes. Indeed, taking into account $[\di,\bbvar{d}]=0$, and repeatedly using Stokes's theorem, we find
\begin{align}
\frac{1}{2\ell}\int_{\mathcal{D}}&{\ast}\left(\di X_\mu\wedge\di X_\nu\right)\wedge\bbvar{d}\nT{1}{\Delta}{^{\mu\nu}}=
-\frac{1}{2\ell}\int_{\mathcal{D}}{\ast}\left(\di X_{[\mu}\wedge X_{\nu]}\right)\wedge\bbvar{d}\big(\di \nT{1}{\Delta}{^{\mu\nu}}\big)=\nonumber\\
&=+\frac{1}{2\ell}\int_{\mathcal{D}}{\ast}\left(\di\big( \bbvar{d}X_{[\mu}\big)\wedge X_{\nu]}\right)\wedge\di \nT{1}{\Delta}{^{\mu\nu}}
=-\frac{1}{2\ell}\int_{\mathcal{D}}{\ast}\left(\bbvar{d}X_{[\mu}\wedge\di X_{\nu]}\right)\wedge\di \nT{1}{\Delta}{^{\mu\nu}}=0,
\end{align}
where we set all terms to zero that vanish thanks to the Einstein equations \eref{o1EOM1} at order $n=1$.\medskip

We are now in the position to turn our attention from the presymplectic potential to the perturbative expansion of  the presymplectic two-form on $\mathcal{D}$. The presymplectic two-form $\Omega_\mathcal{D}$ is the exterior derivative of the presymplectic potential, i.e.,\
\begin{equation}
\Omega_{\mathcal{D}} = \bbvar{d}\Theta_{\mathcal{D}}.
\end{equation}
Going back to the perturbative expansion of the presymplectic potential \eref{ThetaD1}, we obtain
\begin{align}
\Omega_{\mathcal{D}} =& \sum_{i=1}^N\bbvar{d}p^i_\mu\,\bbvar{d}q^\mu_i+\int_{\mathcal{D}}{\ast}\left(\di X_{[\mu}\wedge\di\bbvar{d}X_{\nu]}\right)\wedge\bbvar{d}\nT{2}{\Delta}{^{\mu\nu}}+\nonumber\\
&+\int_{\mathcal{D}}\left[{\ast}\left(\di \big(\bbvar{d}X_{[\mu}\big)\wedge\nT{1}{f}{_{\nu]}}\right)\wedge\bbvar{d}\nT{1}{\Delta}{^{\mu\nu}}+{\ast}\left(\di X_{[\mu}\wedge\bbvar{d}\nT{1}{f}{_{\nu]}}\right)\wedge\bbvar{d}\nT{1}{\Delta}{^{\mu\nu}}\right]+\nonumber\\
&-\frac{1}{2}\oint_{\partial{\mathcal{D}}}\left[\bbvar{d}\ou{S}{\mu}{\nu}\,\ou{\bbvar{m}}{\nu}{\mu}+\frac{1}{2}\ou{S}{\mu}{\nu}\ou{\left[\bbvar{m},\bbvar{m}\right]}{\nu}{\mu}\right]+\mathcal{O}(\ell).\label{Omexpansion1}
\end{align}
where $\ou{S}{\mu}{\nu}$ is the $\mathfrak{so}(1,3)$-valued spin current \eref{Sdef} and we introduced the Maurer-Cartan form in field space, which is dual to it, i.e.,\
\begin{equation}
\bbvar{m} = \left(\bbvar{d}\Lambda^{-1}\right)\Lambda,\qquad \bbvar{d}\bbvar{m}=\frac{1}{2}\left[\bbvar{m},\bbvar{m}\right].
\end{equation}
The bulk terms can be neatly reorganized. First of all, we introduce a field-space covariant derivative $\bbvar{D}$ given by the ordinary exterior derivative $\bbvar{d}$ on field space shifted by a diffeomorphism,\footnote{The connection is flat, i.e.,\ $\bbvar{D}^2=0.$ The geometry of such field-space connections for gauge theories and gravity is discussed in great detail in \cite{Gomes:2016mwl}.}
\begin{align}
\bbvar{D}\nT{n}{f}{^{\mu}} & = \bbvar{d}\nT{n}{f}{^{\mu}} - \mathcal{L}_{\bbvar{X}}\nT{n}{f}{^{\mu}},\\
\bbvar{D}\nT{n}{\Delta}{^{\mu\nu}} & = \bbvar{d}\nT{n}{\Delta}{^{\mu\nu}} - \mathcal{L}_{\bbvar{X}}\nT{n}{\Delta}{^{\mu\nu}}.
\end{align}
Here, $\bbvar{X}^a$ is a vector-valued one-form on field space, explicitly,
\begin{align}
&\bbvar{X}^a  = \left[\frac{\partial}{\partial X^\mu}\right]^a\bbvar{d}X^\mu,%\\
\end{align}
and $\mathcal{L}$ is the Lie derivative satisfying $\mathcal{L}_\xi(\cdot)  = \xi\hook\left(\di\cdot\right)+\di\left(\xi\hook\cdot\right)$ for all $\xi^a\in\Gamma(T\mathcal{M})$. Next, we reorganize the various terms. Consider first the second term in \eref{Omexpansion1}. Taking into account the linearized Einstein equations at order $n=2$, we can rewrite it as
\begin{align}\nonumber
\small
\int_{\mathcal{D}}{\ast}\left(\right.&\di\left. X_{[\mu}\wedge\di\bbvar{d}X_{\nu]}\right)\wedge\bbvar{d}\nT{2}{\Delta}{^{\mu\nu}}=\int_{\mathcal{D}}\di X_{[\mu}\wedge\di\bbvar{d}X_{\nu]}\wedge\bbvar{d}\big({\ast}\nT{2}{\Delta}{^{\mu\nu}}\big)\\\nonumber
=&\oint_{\partial\mathcal{D}}\bbvar{d}X_{[\mu}\di X_{\nu]}\wedge\bbvar{d}\big({\ast}\nT{2}{\Delta}{^{\mu\nu}}\big)+\int_{\mathcal{D}}\bbvar{d}X_{[\mu}\di X_{\nu]}\wedge\bbvar{d}\di\big({\ast}\nT{2}{\Delta}{^{\mu\nu}}\big)\\\nonumber
=&\oint_{\partial\mathcal{D}}\bbvar{d}X_{[\mu}\di X_{\nu]}\wedge\bbvar{d}\big({\ast}\nT{2}{\Delta}{^{\mu\nu}}\big)-\int_{\mathcal{D}}\bbvar{d}X_{[\mu}\di \big(\bbvar{d}X_{\nu]}\big)\wedge\di\big({\ast}\nT{2}{\Delta}{^{\mu\nu}}\big)+\int_{\mathcal{D}}\bbvar{d}X_{\mu}\bbvar{d}\left({\ast}\di\big(\nT{2}{\Delta}{^{\mu}_{\nu}}\big)\wedge\di X^\nu\right)\\\nonumber
=&\oint_{\partial\mathcal{D}}\bbvar{d}X_{[\mu}\di X_{\nu]}\wedge\bbvar{d}\big({\ast}\nT{2}{\Delta}{^{\mu\nu}}\big)-\frac{1}{2}\oint_{\partial\mathcal{D}}\bbvar{d}X_{[\mu} \bbvar{d}X_{\nu]}\wedge\di\big({\ast}\nT{2}{\Delta}{^{\mu\nu}}\big)+\int_{\mathcal{D}}\bbvar{d}X_{\mu}\bbvar{d}\left(\nT{2}{T}{^\mu}+\nT{2}{t}{^\mu}\right)\\\nonumber
=&\oint_{\partial\mathcal{D}}\bbvar{d}\left(\big({\ast}\nT{2}{\Delta}{_{\mu\nu}}\big)\wedge\di X^\nu\right)\,\bbvar{d}X^{\mu}+\sum_{i=1}^N\bbvar{d}p^i_\mu\,\bbvar{d}X^\mu\Big|_{\gamma_i\cap\mathcal{D}}+\int_{\mathcal{D}}\bbvar{d}{X}_\mu\,\bbvar{d}\nT{2}{t}{^\mu}
\end{align}
We insert this expression back into the presymplectic structure and identify three separate terms at order $\mathcal{O}(\ell^0)$ or lower. Let us first list them and then comment on their physical significance in the next subsection. We have
\begin{equation}
\Omega_{\mathcal{D}} = \Omega_{\mathcal{D}}^{\mtext{matter}} + \Omega_{\mathcal{D}}^{\mtext{rad}} +\Omega_{\partial\mathcal{D}}+\mathcal{O}(\ell).
\end{equation}
The first term is the matter contribution:
\begin{empheq}[box=\mygraybox]{align}
\Omega_{\mathcal{D}}^{\mtext{matter}} = \sum_{i=1}^N\bbvar{d}p^i_\mu\,\big(\bbvar{d}q^\mu_i+\bbvar{d}X^\mu\big|_{\gamma_i\cap\mathcal{D}}\big)\equiv
\sum_{i=1}^N\bbvar{d}p^i_\mu\,\bbvar{D}q_i^\mu.\label{Ommatter}
\end{empheq}
The second term is the radiative part of the presymplectic potential:
\begin{align}
\Omega_{\mathcal{D}}^{\mtext{rad}}  = \int_{\mathcal{D}}\left[{\ast}\left(\di \big(\bbvar{d}X_{[\mu}\big)\wedge\nT{1}{f}{_{\nu]}}\right)\wedge\bbvar{d}\nT{1}{\Delta}{^{\mu\nu}}+{\ast}\left(\di X_{[\mu}\wedge\bbvar{d}\nT{1}{f}{_{\nu]}}\right)\wedge\bbvar{d}\nT{1}{\Delta}{^{\mu\nu}}%\right.\\
+\bbvar{d}{X}_\mu\,\bbvar{d}\nT{2}{t}{^\mu}\right],\label{Omradintermediate}
\end{align}
with $\nT{2}{t}{^\mu}$ denoting the vector-valued three-form \eref{Tdef1}. A straightforward algebraic manipulation (see \hyperref[appdxB]{Appendix A}) simplifies this expression, which can be cast into
\begin{empheq}[box=\mygraybox]{equation}
\Omega_{\mathcal{D}}^{\mtext{rad}}  = \int_{\mathcal{D}}{\ast}\left(\di X_{[\mu}\wedge\bbvar{D}\nT{1}{f}{_{\nu]}}\right)\wedge\bbvar{D}\nT{1}{\Delta}{^{\mu\nu}}.\label{Omrad2}
\end{empheq}
Finally, there is the boundary term
\begin{empheq}[box=\mygraybox]{equation}
\Omega_{\partial\mathcal{D}}=\oint_{\partial\mathcal{D}}\bbvar{d}P_\mu\,\bbvar{d}X^{\mu}-\frac{1}{2}\oint_{\partial{\mathcal{D}}}\left[\bbvar{d}\ou{S}{\mu}{\nu}\,\ou{\bbvar{m}}{\nu}{\mu}+\frac{1}{2}\ou{S}{\mu}{\nu}\ou{\left[\bbvar{m},\bbvar{m}\right]}{\nu}{\mu}\right],\label{bndryOm}
\end{empheq}
where $\ou{S}{\mu}{\nu}$ is the $\mathfrak{so}(1,3)$ spin current \eref{Sdef} and $P_\mu$ is the local momentum
\begin{equation}
P_\mu=\varphi^\ast_{\mathcal{C}}\left(\big({\ast}\nT{2}{\Delta}{_{\mu\nu}}\big)\wedge\di X^\nu\right).\label{localmomentum}
\end{equation}

\paragraph{- Summary} Before we turn to the interpretation of these terms, let us briefly summarize what we have done so far. Starting from general relativity in terms of tetrads, we performed an expansion in terms of the coupling constant $\ell = \sqrt{8\pi G}$ up to second order. We considered  gravity coupled to $N$ massive point particles in a bounded region $\mathcal{D}$ and derived the perturbative expansion of the presymplectic two-form $\Omega_\mathcal{D}$ on phase space. Assuming that the region is big enough such that  gravitational radiation $^{(1)}f$ does not reach the boundary and using the linearized equations of motion, we simplified the expression to obtain three terms, corresponding to matter modes \eqref{Ommatter}, radiation modes \eqref{Omrad2}, and additional boundary modes \eqref{bndryOm}.

\subsection{Modes of gravity}
 
\noindent Let us now explain the physical interpretation of the various contributions to the presymplectic two-form $\Omega_\mathcal{D}$. First of all, we have the contribution from the matter degrees of freedom \eref{Ommatter}. This contribution describes the phase space of the point particles. The particles' states (points on phase space) are labeled by the particles' position and their four-momentum $p_\mu^i$ on the $t=\mathrm{const}$ initial surface $\Sigma$. Notice that the variation of the particles' position is \emph{dressed} by a coordinate transformation replacing the differential $\bbvar{d}q^\mu_i$ by $\bbvar{D}q^\mu_i$, see \eref{Ommatter}. 
This dressing ensures that the symplectic structure of the point particles is invariant under simultaneous diffeomorphisms $\varphi\in\mathrm{Diff}(\mathcal{M}:\mathcal{M})$ transforming both tetrads and trajectories into $e^\alpha\rightarrow\varphi^{\ast}e^\alpha$ and $\gamma_i\rightarrow \varphi^{-1}\circ\gamma_i$. As the coordinate functions transform as scalar fields, i.e.,\ $X^\mu\rightarrow X^\mu\circ\varphi$, the \emph{dressed} differential $\bbvar{D}q^\mu_i=\bbvar{d}q_i^\mu+\bbvar{d}X^\mu|_{\gamma_i}=\bbvar{d}(X^\mu\circ\gamma_i)$ is invariant under such combined transformations of the trajectories and the fields in spacetime.\medskip

Next, there are the radiative modes \eref{Omrad2}. Variations of the first-order radiative perturbations $\nT{1}{f}{^\mu}$ and $\nT{1}{\Delta}{^\mu_\nu}$ appear in the presymplectic two-form only through the covariant field differential $\bbvar{D} = \bbvar{d}-\mathcal{L}_{\bbvar{X}}$.  
Going back to the mode expansion \eref{modexpns}, we find
\begin{equation}
\bbvar{D}\nT{1}{f}{^{\mu}} = \frac{1}{{(2\pi)}^{\frac{3}{2}}}\int \frac{d^3 k}{2|\vec{k}|} \Big(m_\mu m_\nu\,(\bbvar{d}a_+)(\vec{k})\E^{\I k_\mu X^\mu}+\bar{m}_\mu \bar{m}_\nu\,(\bbvar{d}a_-)(\vec{k})\E^{\I k_\mu X^\mu}+\CC\Big)\di X^\nu,
\end{equation}
and similarly for $\nT{1}{\Delta}{^\mu_\nu}$. The symplectic structure of the radiative modes thus depends only on the variations of the true physical degrees of freedom, i.e.,\ the modes $a_\pm(\vec{k})$ that characterize the strength and polarization of gravitational waves in the given gauge. From an operational point of view, the gauge condition that we implicitly used to parametrize the radiation field \eref{modexpns} states that the $X^i=\mathrm{const}$ coordinate lines are trajectories of test particles at rest. These trajectories are geodesics up to terms of order $n=1$ in the perturbative expansion. Gravitational waves will change the relative physical distance between these geodesics, but leave their coordinate distance fixed.\medskip

Finally, we have the boundary contribution \eref{bndryOm}. This is where the otherwise irrelevant gauge modes become physical. There are additional boundary modes $X^\mu$ dual to  diffeomorphisms and  $SO(1,3)$-valued  boundary modes that are dual to the Lorentz charges $S_{\mu\nu}$. These will be the focus of the next section. 

\section{Multi-fingered Reference Frames from Gravitational Boundary Modes}
\label{sec:boundaryModes}
\subsection{Boundary phase space}

\noindent As we have seen in Sec.~\ref{sec:edgemodes}, by the example of electromagnetism, restricting the phase space of a gauge theory to a bounded region leads to additional boundary modes. The $U(1)$ gauge parameter $\lambda$, which is redundant in the description of the physics of the bulk, becomes physical at the boundary. Mathematically, this is realized by the emergence of boundary contributions to the presymplectic two-form.\medskip

In the case of general relativity in the tetrad formulation, we have two such gauge groups. 
The field equations \eref{EEq1} and \eref{EEq2} are invariant under four-dimensional diffeomorphisms and internal $SO(1,3)$ frame rotations. Inside a compact region $\mathcal{D}$ on a Cauchy surface $\Sigma$, it is impossible to distinguish configurations that are connected by small gauge transformations. Any such small gauge transformation can be written as a finite product of diffeomorphisms generated by the exponential map of vector fields $\xi^a_1,\xi^a_2,\dots$ times the exponential of a  local $\mathfrak{so}(1,3)$ gauge element $\ou{\lambda}{\alpha}{\mu}$.  For such transformations to be \emph{small}, both  $\ou{\lambda}{\alpha}{\mu}$ and the vector fields $\xi^a$ have compact support. If we extend them to the boundary, they become physical, generating a \emph{symmetry} that maps a given state into an inequivalent one. The characterization of the physical states of the theory will thus include a description of additional boundary modes. In our setup, where we consider the perturbative expansion of the gravitational field in a neighborhood of Minkowski space, the boundary modes show up as the coordinate fields $X^\mu$ and the $SO(1,3)$ gauge elements $\ou{\Lambda}{\alpha}{\mu}$, arising from the internal Lorentz symmetry. The corresponding contribution to the presymplectic form is given by Eq.~\eqref{bndryOm}.\medskip  

\begin{figure}[h!]
    \centering
    \includegraphics[scale=0.25]{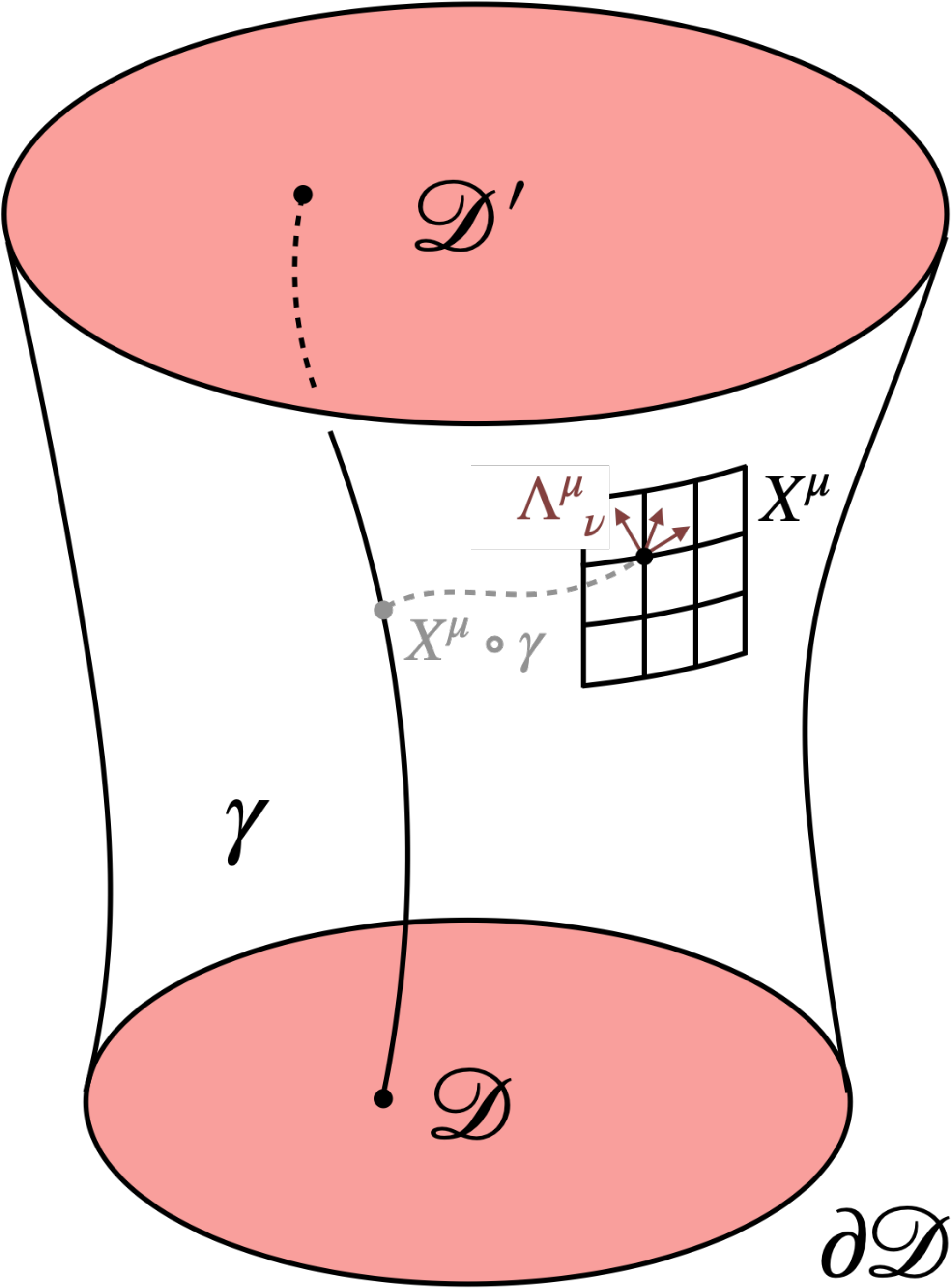}
    \caption{The additional boundary modes for the subregion $\mathcal{D}$: coordinates $X^\mu$ and Lorentz frames $\Lambda^{\alpha}{_{\mu}}$ at each point of the boundary. The coordinate fields serve two purposes: in the bulk, they allow us to define the diffeomorphism-invariant, relative position of the point particle moving along $\gamma$ by expressing the path in terms of the coordinate fields as $X^
    \mu\circ \gamma$ (indicated by the dotted line). In addition, they give physical meaning to the position of the boundary itself by providing an embedding into the flat background spacetime.}
    \label{fig:boundary_modes}
\end{figure}

The coordinate fields $X^\mu$, which are inertial with respect to the background metric, act as physical reference frames with respect to which we can construct gauge-invariant (\emph{dressed}) observables (see Fig.~\ref{fig:boundary_modes}). 
This becomes particularly evident in our setup, where we consider point particles coupled to the gravitational field. In a generally covariant theory, the embedding of a trajectory $\gamma_i$ into the abstract spacetime manifold is meaningless without giving a physical prescription for how to measure the location of the trajectory.  
 The physical reference fields $X^\mu$ allow us to resolve this issue.  The relative position of the trajectory with respect to the coordinate system $X^\mu$ is gauge invariant. To leading order in the perturbation, the particles move along straight lines with respect to the reference frame $X^\mu$, i.e.,\ $(X^\mu\circ\gamma_i)(\tau) = q^\mu_o+m_i^{-1}p^\mu_i\tau$. Along the particles' trajectories, the reference frames become physical and serve as coordinate fields that give physical meaning to the otherwise gauge-dependent position $\gamma_i(\tau)$ of the particles. This relational definition is important, because it allows us to access the gauge-invariant part of the presymplectic two-form of the matter modes through the dressed variation of the particles' positions, $\bbvar{D}q^\mu_i=\bbvar{d}q_i^\mu+\bbvar{d}X^\mu|_{\gamma_i}=\bbvar{d}(X^\mu\circ\gamma_i)$. The differential $\bbvar{D}q^\mu_i$ is gauge invariant, because a diffeomorphism will shift both the trajectories $\gamma_i$ of the particles and the coordinate fields $X^\mu$, in such a way that the dressed quantity $X^\mu\circ\gamma_i$ does not change. In the same way, the fields $\ou{\Lambda}{\alpha}{\mu}$ provide an internal reference frame for the Lorentz group at each point of the boundary. Take, for example, the $SO(1,3)$ curvature tensor $\ou{F}{\alpha}{\beta}$; its $\Lambda$-dressed version $\uo{\Lambda}{\alpha}{\mu}\ou{F}{\alpha}{\beta}\ou{\Lambda}{\beta}{\nu}$ is clearly $SO(1,3)$ gauge invariant.\smallskip

The coordinate fields $X^\mu$ have an additional role. Besides serving as dressing fields that give physical meaning to radiative modes and particle trajectories, they also define the location of the boundary itself. To the zeroth order of the expansion, the coordinate fields $X^\mu$ define an inertial frame of reference, which embeds the abstract boundary of $\mathcal{D}$ into flat Minkowski space $\R^4$. In this sense, the boundary modes $X^\mu$ provide a concrete realization of the embedding fields of \cite{CLP22}. 

 \medskip
Given the presymplectic two-form at the boundary, see \eqref{bndryOm}, we define an extended phase space that contains the conjugate boundary observables that are dual to the boundary reference frame $X^\mu$ and the Lorentz fields $\ou{\Lambda}{\alpha}{\mu}$. First of all,  we have the coordinate fields $X^\mu$ at the boundary, which are dual to the local momentum
\begin{equation}
    P_\mu=\big({\ast}\nT{2}{\Delta}{_{\mu\nu}}\big)\wedge\di X^\nu.\label{Pdef}
\end{equation}
If we send the boundary to infinity, the integral of the local momentum \eref{Pdef} along the asymptotic two-sphere returns the Arnowitt--Deser--Misner (ADM) linear momentum \cite{Arnowitt:1962hi,ADMEnergy,ReggeTeitelboim,Ashtekar:1978zz,AshtekarAsymptcs1980,Beig:1987zz}. The probably simplest way to see this is to use the standard representation of the ADM momentum in terms  of the electric part of the Weyl tensor \cite{Ashtekar:1978zz}. Take hyperbolic coordinates in the vicinity of spacelike infinity,\begin{equation}
X^\mu=\begin{pmatrix}
\rho \sinh(t/\rho) \\ \rho\cosh(t/\rho)\,\sin\vartheta\,\cos\varphi\\
\rho\cosh(t/\rho)\,\sin\vartheta\,\sin\varphi\\
\rho\cosh(t/\rho)\,\cos\vartheta
\end{pmatrix}.
 \end{equation}
Next, we integrate \eref{Pdef} along a $t=\mathrm{const}$ and  $\rho=\mathrm{const}$ surface. Performing a partial integration and taking into account the falloff conditions of the Coulombic fields, we obtain
\begin{align}\nonumber
\lim_{\rho\rightarrow\infty}\oint_{S^2_{\rho,t}}P_\mu=\lim_{\rho\rightarrow\infty}\oint_{S^2_{\rho,t}}\big({\ast}\nT{2}{\Delta}{_{\mu\nu}}\big)\wedge\di X^\nu &=-\lim_{\rho\rightarrow\infty}\oint_{S^2_{\rho,t}}{\ast}\big(\di \nT{2}{\Delta}{_{\mu\nu}}\big)X^\mu =\\
&=\frac{1}{8\pi G}\lim_{\rho\rightarrow\infty}\oint_{S^2_{\rho,t}}d^2\Omega\,\rho^3\,\nT{2}{C}{_{\mu\nu\rho\sigma}}n^\nu n^\rho t^\sigma,\
\end{align}
where $t^\mu=\partial_tX^\mu$ and $n^\mu=\rho^{-1}X^\mu$ are the timelike and spacelike normal vectors to the asymptotic two-sphere, $d^2\Omega$ is the round area element $d^2\Omega=\sin\vartheta\,\di\vartheta\,\di\varphi$, and $\nT{2}{C}{_{\mu\nu\rho\sigma}}n^\nu n^\rho $ denotes the $n=2$ leading order of the electric part of the Weyl tensor, which falls off like $\mathcal{O}(\rho^{-3})$ in the outside region $\bar{\Sigma}=\Sigma-\mathcal{D}$.

Besides the local momentum, there is also a spin current \eref{Sdef} at the boundary, which is dual to the local Lorentz elements $\ou{\Lambda}{\alpha}{\mu}$. In the linearized regime considered here, the local momentum and spin current $P_\mu$ and $S_{\mu\nu}$ depend only linearly on the Coulombic modes $\nT{2}{f}{^\mu}$, which are themselves determined by the effective energy momentum tensor via \eref{2fsourcedby} and \eref{o2EOM2}.\medskip

A few further observations. Using the equations of motion, it is straightforward to see that both the local momentum and angular momentum are conserved. Throughout this work, we assumed that the $n=1$ radiative modes $\nT{1}{f}{^\mu}$ have compact support in $\mathcal{D}$. Together with the second-order Einstein equations \eref{o2EOM1}, this implies immediately that
\begin{equation}
\di P_\mu = 0\;\text{in}\;\bar{\Sigma} =\Sigma-\mathcal{D}.\label{Pconservd}
\end{equation}
In the same way, we have conservation of the local angular momentum, which is the sum of the local orbital angular momentum and the intrinsic spin density \eref{Sdef},
\begin{equation}
J_{\mu\nu} = 2P_{[\mu} X_{\nu]} + S_{\mu\nu}.
\end{equation}
To demonstrate that the local angular momentum  $J_{\mu\nu}$ is conserved in the outside region $\bar{\Sigma}=\Sigma-\mathcal{D}$, see \hyperref[appdxB2]{Appendix B}, where we show that
\begin{equation}
\di J_{\mu\nu} = 0\;\text{in}\;\bar{\Sigma} =\Sigma-\mathcal{D}.
\end{equation}
It is straightforward to see that $J^i[\mathcal{C}]=\tfrac{1}{2}\oint_{\mathcal{C}}\epsilon^{ijk}J_{jk}$ is the ADM angular momentum \cite{Arnowitt:1962hi,ADMEnergy,ReggeTeitelboim} and $K^i[\mathcal{C}]=\oint_{\mathcal{C}}\ou{J}{i}{0}$ returns the Beig--Ó Murchaddha relativistic center of mass (the global boost generator) \cite{Beig:1987zz}.\medskip

Finally, we invert the boundary symplectic structure and obtain the commutation relations,
\begin{align}
\big\{\tilde{P}_\mu(\vec{\zeta}),X^\nu(\vec{\zeta}')\big\}&=\delta^\nu_\mu\tilde{\delta}_{\mathcal{C}}(\vec{\zeta},\vec{\zeta}'),\label{bndryPX}\\
\big\{\tilde{S}_{\mu\nu}(\vec{\zeta}),\ou{\Lambda}{\alpha}{\rho}(\vec{\zeta}')\big\}&=+2\,\eta_{\rho[\mu}\ou{\Lambda}{\alpha}{\nu]}(\vec{\zeta})\tilde{\delta}_{\mathcal{C}}(\vec{\zeta},\vec{\zeta}'),\label{bndrySpin}\\
\big\{\tilde{S}_{\mu\nu}(\vec{\zeta}),\tilde{S}_{\mu'\nu'}(\vec{\zeta}')\big\}&=-4\,\delta_{[\mu}^\rho\delta_{\nu]}^\sigma\eta_{\sigma\sigma'}\delta_{[\mu'}^{\sigma'}\delta_{\nu']}^{\rho'}\tilde{S}_{\rho\rho'}(z)\tilde{\delta}_{\mathcal{C}}(\vec{\zeta},\vec{\zeta}')
\label{bndrySpinSpin},
\end{align}
where  $\vec{\zeta}$ and $\vec{\zeta}'$ label points on the boundary surface $\mathcal{C}$. In here, the local four-momentum $\tilde{P}_\mu$ and the spin current $\tilde{S}_{\mu\nu}$ are now treated as surface densities intrinsic to $\mathcal{C}$. The connection to the  fields in the interior of the manifold is imposed by the momentum and spin-momentum constraints
\begin{align}
\tilde{C}_\mu&:=\tilde{P}_\mu-\tilde{H}_\mu,\label{eq:momentumConstraint}\\
\tilde{C}_{\mu\nu}&:=\tilde{S}_{\mu\nu}-\tilde{E}_{\mu\nu}.\label{eq:spinConstraint}
\end{align}
where the boundary Hamiltonian density $\tilde{H}_\mu$ and the spin surface density $\tilde{S}_{\mu\nu}$ are given by the pull back of the two-forms \eref{Pdef} and \eref{Sdef} to $\mathcal{C}$, i.e.,\
\begin{align}
\tilde{H}_\mu&:=\varphi^\ast_{\mathcal{C}}\Big[\big({\ast}\nT{2}{\Delta}{_{\mu\nu}}\big)\wedge\di X^\nu\Big],\label{H1def}\\
\tilde{E}_{\mu\nu}&:=\varphi^\ast_{\mathcal{C}}\frac{1}{\ell^2}\Big[{\ast}\left(\di X_{[\mu}\wedge\di X_{\nu]}\right)+2\ell^2{\ast}\left(\di X_{[\mu}\wedge\nT{2}{f}{_{\nu]}}\right)\label{H2def}\Big].
\end{align}

We have thus found an extended phase space of bulk and boundary modes and a set of additional constraints \eref{eq:momentumConstraint} and \eref{eq:spinConstraint} that establish the connection between the theory at the boundary and the theory in the interior.

\subsection{Implications for quantum gravity}

\noindent  
The structure of the phase space that has emerged from the discussions in the previous sections has important implications  for the quantum theory. As any full nonlinear and nonperturbative theory of quantum gravity must agree at low energies with the linearized theory, there is no need to select a particular model or approach to quantum gravity. At low energies, foundational questions are shared across different approaches. \medskip
 
Through our preceding analysis, we found a neat separation of the classical state space into matter modes $(p_i^\mu,q^\mu_i)$, radiative modes $a_\pm(\vec{k})$ and boundary modes $X^\mu$ and $\ou{\Lambda}{\alpha}{\mu}$ with corresponding momentum currents $\tilde{P}_\mu$ and $\tilde{S}_{\mu\nu}$.  
A similar decomposition must also happen at the quantum level. At the linearized level,\footnote{At the full nonperturbative level, we would not expect such a factorization as it is no longer obvious how to disentangle radiative modes from the boundary modes emerging from the zeroth-order metric expansion.} we expect a \emph{kinematical} state space that admits a tensor product structure
\begin{equation}
\mathcal{K}_{\mathcal{D}} = \underbrace{\mathcal{H}^{\mtext{matter}}_{\mathcal{D}}\otimes\mathcal{H}^{\mtext{rad}}_{\mathcal{D}}}_{\mathcal{H}_{\mtext{bulk}}}\otimes 
\mathcal{H}^{\mtext{bndry}}_{\partial\mathcal{D}}.
\end{equation}
 The first term is the Hilbert space for the matter degrees of freedom, i.e.,\ the tensor product of the one-particle Hilbert spaces for each constituent system. The second term describes the radiation modes. Inverting the presymplectic two-form at leading order in the coupling constant, we obtain the Poisson bracket between the positive and negative frequency modes $a_{\pm}$ and $\bar{a}_{\pm}$:
\begin{equation}
\big\{a_s(\vec{k}),\bar{a}_{s'}(\vec{k'})\big\}=2\I|\vec{k}|\delta_{ss'}\delta^{(3)}(\vec{k}-\vec{k}')+\mathcal{O}(\ell),\quad s,s'\in\{\pm\}.
\end{equation}
Notice that the $G\rightarrow 0$ decoupling limit removes the nonlinear higher-order terms on the right-hand side. In this limit, the Hilbert space of the radiation modes is generated  from the usual Fock vacuum $|0\rangle$,
\begin{equation}
a_s(\vec{k})\big|0\rangle = 0.
\end{equation}
Gravitons are excited from the vacuum by creation operators $a^\dagger[f]=\sum_s\int d^3k/|\vec{k}|\,f_s(\vec{k})\,a_s^\dagger(\vec{k})$. To obtain an element of $\mathcal{H}_{\mathcal{D}}^{\mtext{rad}}$, the test functions $f_s(\vec{k})$ must be such that their Fourier transform has compact support in $\mathcal{D}$.\medskip 

Besides matter and radiation, there is the boundary Hilbert space $\mathcal{H}_{\partial\mathcal{D}}$, which carries a representation of the canonical commutation relations \eref{bndryPX}, \eref{bndrySpin}, \eref{bndrySpinSpin}. At a formal level, we may choose a functional Schr\"odinger representation, for which the momentum operators %, which are surface densities intrinsic to the two-surface boundary $\mathcal{C}=\partial\mathcal{D}$, 
act as functional derivatives, i.e.,\
\begin{align}
\big(\tilde{P}_\mu(\vec{\zeta}) \Psi_{\mtext{bndry}}\big)[X^\mu,\ou{\Lambda}{\alpha}{\mu}]& = -\I\hbar\frac{\delta}{\delta X^{\mu}(\vec{\zeta})}\Psi_{\mtext{bndry}}[X^\mu,\ou{\Lambda}{\alpha}{\mu}],\\
\big(\tilde{S}_{\mu\nu}(\vec{\zeta})\Psi_{\mtext{bndry}}\big)[X^\mu,\ou{\Lambda}{\alpha}{\mu}]& = 2\I\hbar \Lambda_{\alpha[\mu}(\vec{\zeta})\frac{\delta}{\delta\uo{\Lambda}{\alpha}{\nu]}(\vec{\zeta})}\Psi_{\mtext{bndry}}[X^\mu,\ou{\Lambda}{\alpha}{\mu}]
\end{align}
whereas $X^\mu(\vec{\zeta})$ and $\ou{\Lambda}{\alpha}{\mu}(\vec{\zeta})$ act by multiplication and $\vec{\zeta}=(\vartheta,\varphi)$ are fiducial coordinates at $\partial\mathcal{D}$. So far, the entire construction is formal. The obvious difficulty is that it is not enough to have a representation of the canonical boundary Heisenberg commutation relations. We also need an inner product. The formal infinite-dimensional Lebesgue measure $\mathcal{D}[X^\mu]=\prod_{\vec{\zeta}}d^4X^\mu(\vec{\zeta})$,  gives rise to a vast Hilbert space, which is nonseparable. This is hardly surprising. The algebra of boundary charges is ultralocal. There is no  boundary-intrinsic Hamiltonian $H\sim X^2+P^2$ with respect to which we could built a (separable) boundary Fock space. At first sight, this seems to be a major problem ahead, but we will argue below that the presence of constraints will render this problem rather fictitious. At the level of the kinematical state space, no Fock space structure is required. Only at the level of the physical phase space, do we expect to obtain a separable Hilbert space. This is reminiscent of a similar construction in the context of Loop Quantum Gravity (see \cite{thiemann, LOSTtheorem}), where the constraints are imposed on an auxiliary kinematical Hilbert space, which is nonseparable.   \medskip

Let us now turn our attention to the constraints \eref{H1def} and \eref{H2def} that establish the correlation between the bulk and boundary fields. In quantum theory, there are two ways to look at such constrained systems: solving the constraints before quantizing (reduced quantization) or quantizing first and then imposing the constraints at the level of the wavefunction \cite{Isham92,Ditt04,Giesel:2007wn}. In the latter case, one starts with a kinematical state space with some auxiliary inner product. Physical states are identified by the requirement that the matrix elements of the constraints vanish between any two such physical states. If the constraints are first class, we impose instead the even stronger condition that the constraints annihilate all physical states \cite{HennauxTeitelboim}. Given the specific form of the constraints \eqref{eq:momentumConstraint} and \eqref{eq:spinConstraint} as a difference of a canonical momentum and a complicated function of the bulk configuration variables, the constraints can be given a particularly neat interpretation. They provide a relational evolution for the wavefunction of the modes in the bulk with respect to the boundary modes $X^\mu(\vec{\zeta})$ and $\ou{\Lambda}{\alpha}{\mu}(\vec{\zeta})$. More concretely, consider the quantization of the constraint \eqref{eq:momentumConstraint} on the physical states in the configuration space basis. Assuming, for a moment, that all constraints are first class, we obtain a local  Schr\"odinger equation at the boundary, i.e.,\
\begin{equation}
 \hat{P}_\mu(\vec{\zeta}) \Psi_{phys}[X^\mu,\ou{\Lambda}{\alpha}{\mu}] = -\I\hbar\frac{\delta}{\delta X^\mu(\vec{\zeta}) } \Psi_{phys}[X^\mu,\ou{\Lambda}{\alpha}{\mu}] =  \hat{H}_\mu(\vec{\zeta}) \Psi_{phys}[X^\mu,\ou{\Lambda}{\alpha}{\mu}].\label{SchEq}
 \end{equation}
Here, the dependence of the wavefunction $\Psi_{phys}[X,\Lambda]\in\mathcal{H}_{\mtext{bulk}}\otimes\mathcal{H}_{\partial\mathcal{D}}$ on the bulk degrees of freedom is left implicit. Notice that there are four such equations for every point $\vec{\zeta}$ at the codimension-two boundary $\mathcal{C}=\partial\mathcal{D}$ generating local translations in $X^\mu(\vec{\zeta})$. Such \emph{multi-fingered} evolution equations frequently appear whenever we consider the relational evolution of gravitational observables in terms of material frames of Refs.~\cite{Rovelli_1991,Brown:1994py,Giesel:2007wn}. However, there are two crucial differences to the usual formalism. First, the reference fields are now built from gravity itself,\footnote{By choosing e.g.\ harmonic gauge conditions implicitly used in \eref{modexpns}.} without introducing any matter. Second, the constraints live in one dimension less, at the codimension-two boundary of $\mathcal{D}$, where they describe the relational evolution of the wavefunction in the bulk relative to the coordinate fields $X^\mu(\vec{\zeta})$ at the boundary. There is, however, a subtle difficulty with imposing \eref{SchEq} strongly. It is not at all obvious whether the constraints \eref{eq:momentumConstraint} and \eref{eq:spinConstraint} are first class, i.e.,\ whether the (classical or quantum) algebra of the constraints closes under the bracket.\footnote{Recently, a great deal of attention was given to the idea to define modified brackets with respect to which the algebra closes. In this case, all diffeomorphisms are generated by a corresponding surface charge \cite{Barnich:2011mi,Speranza:2017gxd,Freidel21,CLP22,Ciambelli:2022cfr,Wieland:2021eth,Kabel:2022efn}. These charges, in turn, provide a representation of the algebra of vector fields on phase space. A similar approach, which is based on metriplectic geometry, was introduced recently by two of the authors of this research; see \cite{Kabel:2022efn}.  In the metriplectic framework, the bracket depends on dissipation. If there is dissipation, the charges no longer form a closed algebra, yet their  Hamiltonian flow is still well defined.} 
 In fact, our expectation is that they do not close at finite distance and that there is an anomaly due to cross terms such as $\{\tilde{P}_\mu(\vec{\zeta}),\tilde{H}_\nu(\vec{\zeta}')\}\neq 0$.\medskip

At spacelike infinity, when we send $\rho=\sqrt{X_\mu X^\mu}$ to infinity, this problem disappears due to the falloff conditions---the cross terms go to zero. However, this limit also creates new difficulties. The boundary mode $X^\mu$ diverges as $\rho\rightarrow\infty$. To remove this divergence, we proceed as follows. Split a generic reference frame $\{X^\mu\}$ into a divergent part and a finite shift $Q^\mu(\vec{\zeta})$,
\begin{equation}
X^\mu = \ou{\Omega}{\mu}{\nu}\big(X^\nu_o+Q^\nu(\vec{\zeta})\big)+\mathcal{O}(\rho^{-1}),\label{Xregular}
\end{equation}
where $\vec{\zeta}=(\vartheta,\varphi)$ are coordinates at the asymptotic two-sphere. The divergent part consists of a global  (yet arbitrary) Lorentz transformation $\ou{\Omega}{\mu}{\nu}$ in $SO(1,3)$, i.e.,\ $\di\Omega = 0, \bbvar{d}\Omega\neq 0$, and a fiducial reference frame, which is kept fixed once and for all, i.e.,\ $\bbvar{d}X^\mu_o=0$. The global $SO(1,3)$ transformation is required to have access to the global ADM angular momentum and boost charges.  
The finite translation $Q^\mu({\vec{\zeta}})$, on the other hand, is an arbitrary Lorentz vector-valued function on the asymptotic two-sphere. The resulting IR-regularized variation of the boundary modes is then given by
\begin{equation}
\bbvar{d}X^\mu = -\ou{[\Omega\bbvar{d}\Omega^{-1}]}{\mu}{\nu}X^\nu+\ou{\Omega}{\mu}{\nu}\bbvar{d}Q^\nu+\mathcal{O}(\rho^{-1}).
\end{equation}
For any resulting reference frame $X^\mu$, there is a corresponding metric tensor $\eta_{ab}=\eta_{\mu\nu}\partial_aX^\mu\partial_bX^\nu$. In addition, we also have the fiducial reference metric $\eta^o_{ab}=\eta_{\mu\nu}\partial_aX^\mu_o\partial_bX^\nu_o$. Clearly, both metric tensors are flat. Taking into account the falloff condition for the coordinate transformation \eref{Xregular},  the two line elements can only differ by a tensor which is $\mathcal{O}(\rho^{-1})$. The map from  $X^\mu_o$ to $X^\mu$ defines therefore an asymptotic symmetry, namely an asymptotic BMS transformation \cite{Sachs,Bondi21,AshtekarNullInfinity,Ashtekar:1978zz} in standard Minkowskian coordinates.

The analogous  regularization for the $SO(1,3)$ spin frame of reference is
\begin{equation}
\ou{\Lambda}{\alpha}{\mu}=\Big(\delta^\alpha_\nu+\frac{1}{\rho}\ou{\lambda}{\alpha}{\nu}(\vec{\zeta})+\mathcal{O}(\rho^{-2})\Big)\uo{\Omega}{\mu}{\nu}\equiv\ou{M}{\alpha}{\nu}\uo{\Omega}{\mu}{\nu},\label{Lregular}
\end{equation}
where $\ou{\lambda}{\mu}{\nu}(\vec{\zeta})$ is an $\mathfrak{so}(1,3)$ gauge element intrinsic to the two-surface boundary $\mathcal{C}=\partial\mathcal{D}$. It is  possible to show that, given this specific falloff, the boundary presymplectic two-form \eref{bndryOm} has a finite limit to spacelike infinity.\footnote{The contribution from the spin current still contains a divergent part, but this term is a total derivative due to $\bbvar{d}X^\mu_o=0$. Hence, it creates a canonical transformation. In quantum theory, canonical transformations are often realized as unitary maps, unless they are not, in which case different Hamiltonian formulations of the same dynamical system describe the semiclassical limit of unitarily inequivalent quantum theories. This seems to us, in fact, the main origin for the disagreement between a standard perturbative Fock quantization and nonperturbative approaches such as loop quantum gravity. } The resulting boundary phase space is the cotangent bundle of the BMS group times an ultralocal contribution $[T^\ast \mathfrak{so}(1,3)]^{\mathcal{C}}$. A detailed analysis of this phase space is beyond the scope of this paper and will be presented elsewhere. For the following discussion, the important point is that this construction leaves us with an algebra of constraints that is first class. 

For concreteness, let us restrict our final remarks to the BMS part of the asymptotic symmetries. The case of asymptotic and internal frame rotations \eref{Lregular} follows in complete analogy, but more care is needed because of the IR divergence induced by $\tilde{E}_{\mu\nu}$, which diverges quadratically as $\rho\rightarrow\infty$, see \eref{H2def}.   A generic state in the kinematical Hilbert space $\mathcal{K}$ will now be entangled between the bulk and boundary degrees of freedom. It will be constructed from a quantum superposition of bulk states $\Psi_{\mtext{bulk}}[Q^\mu,\ou{\Omega}{\mu}{\nu})\in\mathcal{H}_{\mtext{bulk}}$ that depend parametrically on the asymptotic BMS frame of reference,\footnote{The notation $\Psi_{\mtext{bulk}}[Q^\mu,\ou{\Omega}{\mu}{\nu})$ indicates that the bulk state depends as a functional on the BMS translation $Q^\mu:S_2\rightarrow\R^4$ and as a ordinary function on $\ou{\Omega}{\mu}{\nu}\in SO(1,3)$.} i.e.,

\begin{equation}
|\Psi\rangle = \int_{[\R^4]^{S_2}}\mathcal{D}[Q^\mu]\int_{SO(1,3)}d\mu_\Omega\,\Psi_{\mtext{bulk}}[Q^\mu,\ou{\Omega}{\mu}{\nu})\otimes|Q^\mu,\ou{\Omega}{\mu}{\nu}\rangle,
\end{equation}
where $d\mu_\Omega$ is the left- and right-invariant measure $d\mu_\Omega=\mathrm{Tr}(\bigwedge^6\Omega^{-1}d\Omega)$ on $SO(1,3)$ and $\mathcal{D}[Q^\mu]$ is a formal Lebsgue measure for the translational subgroup, which is infinite dimensional. At this point, the construction is formal, as we have not specified the Hilbert space inner product between the states $|Q^\mu,\ou{\Omega}{\mu}{\nu}\rangle$. 
Still, we can proceed to discuss the basic structure of the resulting physical Hilbert space. Physical states are those that satisfy a multi-fingered Schr\"odinger equation with respect to asymptotic BMS translations,  
\begin{equation}
\big({{\tilde{C}}_\mu(\vec{\zeta})}\Psi\big)[Q^\mu,\ou{\Omega}{\mu}{\nu}):=\I\hbar\frac{\delta}{\delta Q^\mu(\vec{\zeta})}\Psi[Q^\mu,\ou{\Omega}{\mu}{\nu})+\big({{\tilde{H}}_\mu(\vec{\zeta})}\Psi\big)[Q^\mu,\ou{\Omega}{\mu}{\nu})=0,\label{SchEq1}
\end{equation}
where $\tilde{H}_\mu$ is the momentum density \eref{H1def}.  In the same way, there will also be a constraint for the global Lorentz charge, 
\begin{equation}
\I\hbar\frac{\di}{\di t}\bigg|_{t=0}\Psi\big[\exp(-t\omega)\triangleright Q^\mu,\ou{\Omega}{\mu}{\nu}\triangleleft\exp(t\omega)\big) =- \frac{1}{2}\big(\omega^{\mu\nu}J^o_{\mu\nu}\Psi\big)[Q^\mu,\ou{\Omega}{\mu}{\nu}),\label{SchEq2}
\end{equation}
where $\Lambda\triangleright Q^\mu=\ou{\Lambda}{\mu}{\nu}Q^\nu$
 is a left action and $\ou{\Omega}{\mu}{\nu}\triangleleft \Lambda=\ou{\Omega}{\mu}{\rho}\ou{\Lambda}{\rho}{\nu}$ is a right action for any $\ou{\Lambda}{\mu}{\nu}\in SO(1,3)$. The corresponding Hamiltonian is $J^o_{\mu\nu}$, whose space-space and spacetime components are the ADM angular momentum and Beig--Ó Murchaddha center of mass. In our representation,
\begin{equation}
J_{\mu\nu}^o=\frac{1}{2}\oint_{i^o}d^2\Omega\,\tilde{J}_{\mu\nu}^o,
\end{equation}
where $\tilde{J}_{\mu\nu}^o$ is the two-surface density
\begin{equation}
\tilde{J}_{\mu\nu}^o = 2\tilde{H}_{[\mu}X^o_{\nu]}+2\,\varphi^\ast_{i^o}\big(\di X^o_{[\mu}\wedge\nT{2}{f}{_{\nu]}}\big).
\end{equation}

At the full nonperturbative and background-independent level, the dynamics will be encoded into a generalized projector $\boldsymbol{P}$ that maps kinematical states $\Psi\in\mathcal{K}$ into physical states; see \cite{rovelli,alexreview, thiemann, lqcspinfoam1}. As we see now, a  remnant of this generalized projector survives the decoupling limit that we have studied in this paper. The projector onto physical states is the formal delta distribution on the constraints
\begin{equation}
\boldsymbol{P} = \int_{[\R^4]^{S_2}}\mathcal{D}[N]\int_{SO(1,3)} d\mu_{\exp(\omega)}\,\exp\left(\frac{\I}{\hbar}\oint_{S_2}N^\mu\tilde{C}_\mu\right)\exp\left(\frac{\I}{2\hbar} \omega^{\mu\nu}\tilde{C}_{\mu\nu}\right).\label{Prjectr}
\end{equation}

Formally, the physical inner product is defined by the matrix elements of the generalized projector on the kinematical state space
\begin{equation}
\langle\Psi|\Psi'\rangle_{\mtext{phys}} = \langle\Psi|\boldsymbol{P} \Psi'\rangle_{\mtext{kin}}.
\end{equation}
Consider now two such kinematical states. Take, for example, a simple product state between the bulk degrees of freedom and a quantum reference frame peaked perfectly on the configurations $Q_0$ and $\Omega_0$,
\begin{align}
\Psi & = \Psi[Q_0,\Omega_0]\otimes|Q_0,\Omega_0\rangle,
\end{align}
and an entangled state
\begin{align}
\Phi & = \sum_i \Phi_i[Q_i,\Omega_i]\otimes|Q_i,\Omega_i\rangle,
\end{align}
where the reference frame is in a quantum superposition of several such configurations. Assume a delta function normalization for the kinematical inner product, i.e.,\ $\langle Q,\Omega|Q',\Omega'\rangle=\delta(Q-Q')\delta(\Omega-\Omega')$. Under this assumption, the formal matrix elements of the projector \eref{Prjectr} become 
\begin{equation}
\langle\Psi,\Phi\rangle_{\mtext{phys}}=\sum_i\big\langle\Psi[Q_0,\Omega_0]\big|\big(U^{\mathrm{BMS}}_{i\to 0}\Phi_i\big)[Q_i,\Omega_i]\big\rangle_{\mtext{bulk}},\label{QRFtrafo}
\end{equation}
where
\begin{align}
    U^{\mathrm{BMS}}_{i\rightarrow 0} &= \langle Q_0, \Omega_0| \boldsymbol{P} |Q_i, \Omega_i\rangle \label{eq:BMStrafo}
\end{align}
defines a representation of the BMS transformation on the bulk Hilbert space that maps the BMS frame $\{Q_i,\Omega_i\}$ onto $\{Q_0,\Omega_0\}$. This viewpoint resonates with approaches to quantum gravity such as loop quantum gravity  \cite{rovelli, alexreview,Thiemann:2013lka}, where there is a kinematical boundary Hilbert and an amplitude map that is defined via a generalized projector $\boldsymbol{P}$. This projector can act on widely different boundary states. Following this logic, Eq.~\eqref{eq:BMStrafo} can be used to construct the corresponding matrix elements of BMS transformations. {The entire structure is, of course, formal, but, even at the formal level,} the emerging geometry is clear. The kinematical state space $\mathcal{K}$ is a principal fiber bundle $(\mathcal{K},\mathcal{H}_{\mtext{phys}},\mathrm{BMS},\boldsymbol{P})$, the base space is the physical Hilbert space $(\mathcal{H}_{\mtext{phys}},\langle\cdot|\cdot\rangle_{\mtext{phys}})$, which is isomorphic---but not in a canonical way---to the bulk Hilbert space, i.e.,\ the tensor product between the relativistic Hilbert space of matter and the Fock space of the radiation modes. The standard fibers are the BMS frames $(Q^\mu,\ou{\Omega}{\mu}{\nu})$ on which the BMS group acts freely; the projector onto the base manifold is  $\boldsymbol{P}$, which defines the physical inner product. Consider then a specific model of quantum gravity, defined by a projector $\boldsymbol{P}$, see, e.g.,\ \cite{rovelli, alexreview,Thiemann:2013lka} and calculate its matrix elements to infer, via Eq.~\eqref{eq:BMStrafo}. If we take the matrix elements of $\boldsymbol{P}$ between asymptotic boundary states, it would be possible, in principle, to define via \eref{eq:BMStrafo} the matrix elements of asymptotic symmetries at the full nonperturbative level. Our long-term goal is to understand in this way the IR structure of quantum gravity from within a full nonperturbative approach.

Furthermore, the sum $\sum_i (U_{i\to 0}^{\mathrm{BMS}}\Phi_i)[Q_i,\Omega_i]$ in Eq.~\eqref{QRFtrafo} implements what is called a \emph{quantum-controlled} reference frame transformation, transforming the quantum state of the bulk degrees of freedom with a different unitary map $U_{i\to0}^\mathrm{BMS}$ that depends on the configuration of the reference frame $(Q_i, \Omega_i)$. It thus constitutes a quantum reference frame transformation in the sense of e.g.,\ \cite{Giacomini_2019, vanrietvelde2018change, hoehn2019trinity}. To make this connection more explicit and to get a better intuition of this transformation in general, let us restrict ourselves to a change between two reference frames that differ only by translations. That is, consider states for which $\Omega_i = \Omega_0$. In this case, the expression simplifies to

\begin{equation}
    U_{i\to 0} = e^{-\frac{i}{\hbar}\oint_{S^2}(Q_i^\mu-Q_0^\mu)\tilde{H}_\mu},\label{eq:translationUnitary}
\end{equation}

\noindent implementing a pointwise translation by $Q_i^\mu-Q_0^\mu$ at each point of the boundary $S_2$, whose action on the bulk degrees of freedom is defined by $\tilde{H}_\mu$, which we assume is self-adjoint. 
The operator given in Eq.~\eqref{eq:translationUnitary} generalizes the existing quantum reference frame transformations for the translation group \cite{Giacomini_2019, vanrietvelde2018change} and Lorentz group \cite{Apadula2022} in two ways: by going beyond global translations to pointwise translations at the boundary and by identifying their action on the gravitational degrees of freedom within the linearized regime.
Moreover, these transformations provide a further step towards constructing general quantum-controlled coordinate transformations and their action on the geometry, as considered abstractly in \cite{delaHamette2022}. In this respect, they are complementary to the transformations considered in \cite{giacomini2021einsteins, giacomini22einsteins}, which allow one to change into the local inertial frame of a quantum particle, and \cite{delaHamette2021falling}, which implement general coordinate transformations but rely on an auxiliary system to identify the different branches of the superposition and do not include the action on the gravitational field.\medskip

\section{Outlook and Discussion}
\label{sec:conclusion}

\noindent Physics describes correlations among observable facts in terms of fundamental laws of nature. These laws frequently make use of auxiliary elements that are not themselves observable.  
In electrodynamics, the charge current $j^\mu$ couples to the electromagnetic field via the vector potential $A_\mu$. In linearized gravity, the energy momentum tensor $T^{\mu\nu}$ couples to the gravitational field via the metric perturbation $f_{\mu\nu}$. Both $A_\mu$ and $f_{\mu\nu}$ are not directly observable. They are the channels through which component systems interact. The most elementary way to access these otherwise invisible channels appears when we cut the gauge system along a boundary. At the surface that separates the two component parts, gauge invariance is broken. The then necessary boundary modes carry a representation of the fundamental gauge group.\medskip    

The purpose of this study is to provide a complete characterization of gravitational boundary modes in the simplified setting of linearized gravity in a finite domain. Our discussion started out in \hyperref[sec:edgemodes]{Sec.~2} with a brief overview of the situation in electrodynamics. In  \hyperref[sec:ModesOfGravity]{Sec.~3},  we considered the perturbative expansion of the Einstein-Cartan equations coupled to a system of $N$ point particles. At zeroth order in the coupling constant $\ell=\sqrt{8\pi G}$, all propagating degrees of freedom, gravitational waves and matter, are washed away. At this order, we are left with empty Minkowski space. We saw that there are infinitely many ways to parametrize Minkowski space in terms of inertial coordinates $\{X^\mu:\mathcal{D}\rightarrow\R^4\}$ and local Lorentz frames $\{\ou{\Lambda}{\alpha}{\mu}:\mathcal{D}\rightarrow SO(1,3)\}$. Are these different representations of Minkowski space all gauge equivalent? They are not. They are not, because there is a boundary $\partial\mathcal{D}=\mathcal{C}$, which breaks gauge invariance. At the boundary, the coordinate fields $X^\mu$ are dual to the local momentum \eqref{localmomentum} while the Lorentz frame $\ou{\Lambda}{\alpha}{\mu}$ is conjugate to the $\mathfrak{so}(1,3)$-valued boundary area two-form \eref{Sdef}. For any finite domain $\mathcal{D}$, these boundary currents have a non-vanishing limit as $G\rightarrow0$. 
 In this sense, even empty space has infinitely many physical boundary modes, which provide reference frames for the gauge symmetries of gravity.

 Next, we studied the perturbative expansion. At order $n=1$, we have the free radiation field. At the next to leading order, we have the Coulombic fields, which are sourced by the effective energy-momentum tensor. Assuming no incoming radiation from past infinity at order $n=2$, the second-order perturbation $\nT{2}{f}{^\mu}$ is fully determined by matter and radiation. Upon inserting the perturbative expansion back into the presymplectic two-form, we identified two terms. The symplectic structure is the sum of a bulk contribution and a boundary contribution. At lowest order in the expansion, the bulk symplectic structure depends only on the radiation field and the states of matter. The boundary symplectic structure is where the \emph{Coulombic} part of the gravitational field enters. It enters the boundary term through the local momentum $P_\mu$ and the boundary spin current $S_{\mu\nu}$; see Eqs.~\eref{Pdef} and \eref{Sdef}. The local momentum  $P_\mu(\vec{\zeta})$ and the  spin current $S_{\mu\nu}(\vec{\zeta})$ are conjugate to the local reference frame $X^\mu(\vec{\zeta})$ and the internal spin frame of reference $\ou{\Lambda}{\alpha}{\mu}(\vec{\zeta})$ at the boundary. Let us also stress that in our analysis the boundary modes emerge already from the zeroth-order flat metric. In other approaches, the boundary modes are introduced by hand. For example, in  \cite{CL21,CLP22}, they arise as embedding fields that give physical meaning to the boundary of the subregion. Another approach was discussed in \cite{CarrozzaHoehn21,CEH22}, where they emerge from within the full nonperturbative theory through post-selection. In \cite{DF16}, edge modes are added to render the total presymplectic two-form gauge invariant.  In a similar way, it is possible to extend the phase space such  that the Komar charges form a closed algebra under the Poisson bracket \cite{CL21,CLP22,Freidel21}.  
 In our approach, nothing is added. The boundary modes $(X^\mu, \ou{\Lambda}{\alpha}{\mu})$ are the fields that parametrize the flat metric, which is treated as a $q$ number as well.

Finally, we considered the quantization of the phase space.  
In a full nonperturbative theory of quantum gravity, all geometry will be subject to quantum fluctuations. In \hyperref[sec:boundaryModes]{Sec.~4}, we saw how to realize the low-energy regime of such a quantum geometry within perturbative gravity. Besides matter and radiation, there are the boundary contributions to phase space, which need to be taken into account in the quantum theory, too.  
Quantization then proceeds in a rather conventional way. Ignoring higher orders in the coupling constant, the positive and negative frequency modes of the radiation field turn into creation and annihilation operators of a free field on Minkowski space. Since all elements of phase space ought to be quantum, we also need a quantum representation of the reference frames $(X^\mu,\ou{\Lambda}{\alpha}{\mu})$. At the kinematical level, the boundary charges $(P_\mu(\vec{\zeta}),S_{\mu\nu}(\vec{\zeta}))$ are fields intrinsic to the boundary, dual to the boundary reference frames $(X^\mu(\vec{\zeta}),\ou{\Lambda}{\alpha}{\mu}(\vec{\zeta}))$. Physical states lie on the  hypersurface defined by the momentum and spin-momentum constraints, i.e.,\  \eref{eq:momentumConstraint} and \eref{eq:spinConstraint}. These boundary constraints are what remains of the Wheeler-DeWitt scalar, vector and Gauss constraint of the Einstein-Cartan Hamiltonian system in the linearized regime. The constraints  lead to a multi-fingered Schrödinger equation, which describes a relational evolution of the quantum state with respect to the reference fields at the boundary. Since the reference fields are part of the phase space, different subregions are now realized as different configurations of the bulk plus boundary phase space. Depending on the boundary and falloff conditions at $\partial{\mathcal{D}}$, the system may be open or closed. If it is closed, there exists a Hamiltonian that drives the time evolution with respect to the reference frames at the boundary.\footnote{This happens, for example, when we blow up the region $\mathcal{D}$ into a complete Cauchy surface.} If the system is open, we can work with a modified bracket that takes into account dissipation \cite{Kabel:2022efn}.

At finite distance, a further  subtlety arises. We expect that the charges are not conserved under the Poisson bracket.  To resolve this issue of anomalies at finite distance, we considered the $\rho\rightarrow\infty$ limit. On phase space, this limit is subtle, because $X^\mu(\vec{\zeta})$ diverges as $\rho=\sqrt{X_\mu X^\mu}$ goes to infinity.  To remove this infrared divergence, we made the following observation. Although $X^\mu(\vec{\zeta)}$ may diverge, a variation $\delta X^\mu(\vec{\zeta)}\equiv\delta Q^\mu(\vec{\zeta})$ may still be finite. In this way, we found a rather immediate derivation of the asymptotic BMS translations and their dual charges $P_\mu(\vec{\zeta})$. Given the conservation law \eref{Pconservd} for the local momentum and taking into account the asymptotic expansion of the Green's function near spacelike infinity, it is easy to check that $P_\mu$ is, in fact, conserved along the $\partial_tX^\mu$ generators of the asymptotic hyperboloid $\rho=\sqrt{X_\mu X^\mu}\rightarrow\infty.$ Thus, we obtain a familiar result: there are an infinite number of angle-dependent conserved charges at spacelike infinity \cite{Strominger:2013jfa}. At the quantum level, these conservation laws turn into  a multi-fingered Schr\"odinger equation. The  quantum states in the bulk  evolve with respect to a BMS quantum reference frame at the boundary. Thus, our analysis extends the existing work on quantum reference frames (QRF) to asymptotic symmetry groups, which are infinite dimensional. Moreover, it contributes to a growing body of recent results that demonstrate the close connection of QRFs to gravitational physics, and boundary modes, e.g.,\ \cite{castroruiz2019time,giacomini2021einsteins,cepollaro2021quantum,delaHamette2021falling,Foo2022, Kabel2022conformal, CEH22}. In particular, QRF transformations that realize global translations and rotations \cite{delaHamette2021falling}, global Lorentz transformations \cite{Apadula2022}, and local coordinate transformations \cite{giacomini2021einsteins,giacomini22einsteins} have been developed and applied to gravitational physics. Our work strengthens the connection between these different fields. First of all, we established the identity between boundary modes and QRFs explicitly in the linearized regime. Moreover, we extended the existing framework by providing the QRF transformations for the BMS group, which amount to pointwise translations and Lorentz rotations at the boundary.\medskip

Let us close with a few final remarks on how our results relate to nonperturbative approaches to quantum gravity. We mentioned above that even an empty region of space has infinitely many boundary modes. This observation is crucial. In nonperturbative approaches such as Spinfoams, Loop Quantum Gravity, Group Field Theory and Quantum Regge Calculus \cite{rovelli, thiemann, alexreview,Oriti09,lollreview,Loll:2019rdj}, a quantum spacetime consists of a discrete number of flat building blocks that are glued together with deficit angles around codimension-two corners \cite{flppdspinfoam,Freidel:1998pt,alexreview,contphas,Bianchi:2009tj,Wieland:2016exy,Wieland:2013cr}. A smooth  geometry emerges at large scales from the coarse-grained behavior of many such building blocks \cite{Oriti14,Dittrich14,Asante:2022dnj}. By imposing flatness inside, all physical degrees of freedom are the boundary modes at the now internal two-surface defects. There is no radiation inside the fundamental building blocks; only the boundary modes are quantized. In perturbative gravity, one takes the opposite extreme---the radiation modes are quantum but the boundary modes are classical. Here, we steer a middle course. There is only a single boundary and we consider the quantum theory of both the bulk and boundary modes in the linearized regime.

\section*{Acknowledgements}
\noindent We acknowledge financial support by the Austrian Science Fund (FWF) through BeyondC (Grant No. F7103-N48). V.K. acknowledges support through a DOC Fellowship of the Austrian Academy of Sciences. This publication was made possible through the support of Grants ID No.~61466 and ID No.~62312 from the John Templeton Foundation, as part of The Quantum Information Structure of Spacetime (QISS) Project (qiss.fr). The opinions expressed in this publication are those of the authors and do not necessarily reflect the views of the John Templeton Foundation.

\bibliography{bibliography}

\providecommand{\href}[2]{#2}\begingroup\raggedright\begin{thebibliography}{10}

\bibitem{Weinberg:1964ew}
S.~Weinberg, ``{Photons and Gravitons in $S$-Matrix Theory: Derivation of
  Charge Conservation and Equality of Gravitational and Inertial Mass},'' {\em
  Phys. Rev.} {\bf 135} (1964) B1049--B1056.

\bibitem{DF16}
W.~{Donnelly} and L.~{Freidel}, ``{Local subsystems in gauge theory and
  gravity},'' {\em Journal of High Energy Physics} {\bf 2016} (Sept., 2016)
  102, \href{http://arXiv.org/abs/1601.04744}{{\tt arXiv:1601.04744}}.

\bibitem{Speranza18}
A.~J. {Speranza}, ``{Local phase space and edge modes for
  diffeomorphism-invariant theories},'' {\em Journal of High Energy Physics}
  {\bf 2018} (Feb., 2018) 21, \href{http://arXiv.org/abs/1706.05061}{{\tt
  arXiv:1706.05061}}.

\bibitem{CL21}
L.~Ciambelli and R.~G. Leigh, ``{Isolated surfaces and symmetries of
  gravity},'' {\em Physical Review D} {\bf 104} (2021), no.~4, 046005,
  \href{http://arXiv.org/abs/2104.07643}{{\tt arXiv:2104.07643}}.

\bibitem{CLP22}
L.~{Ciambelli}, R.~G. {Leigh}, and P.-C. {Pai}, ``{Embeddings and Integrable
  Charges for Extended Corner Symmetry},'' {\em Physical Review D} {\bf 128}
  (Apr., 2022) 171302, \href{http://arXiv.org/abs/2111.13181}{{\tt
  arXiv:2111.13181}}.

\bibitem{Freidel21}
L.~Freidel, ``{A canonical bracket for open gravitational system},'' Nov.,
  2021.
\newblock \href{https://arxiv.org/abs/2111.14747}{arXiv:2111.14747}.

\bibitem{Freidel:2020xyx}
L.~Freidel, M.~Geiller, and D.~Pranzetti, ``{Edge modes of gravity. Part I.
  Corner potentials and charges},'' {\em JHEP} {\bf 11} (2020) 026,
  \href{http://arXiv.org/abs/2006.12527}{{\tt arXiv:2006.12527}}.

\bibitem{FOPS21}
L.~Freidel, R.~Oliveri, D.~Pranzetti, and S.~Speziale, ``{Extended corner
  symmetry, charge bracket and Einstein\textquoteright{}s equations},'' {\em
  JHEP} {\bf 09} (2021) 083, \href{http://arXiv.org/abs/2104.12881}{{\tt
  arXiv:2104.12881}}.

\bibitem{Wieland21}
W.~{Wieland}, ``{Null infinity as an open Hamiltonian system},'' {\em Journal
  of High Energy Physics} {\bf 2021} (Apr., 2021) 95,
  \href{http://arXiv.org/abs/2012.01889}{{\tt arXiv:2012.01889}}.

\bibitem{Wieland:2021vef}
W.~Wieland, ``{Gravitational SL(2, \ensuremath{\mathbb{R}}) algebra on the
  light cone},'' {\em JHEP} {\bf 07} (2021) 057,
  \href{http://arXiv.org/abs/2104.05803}{{\tt arXiv:2104.05803}}.

\bibitem{Wieland:2017cmf}
W.~Wieland, ``{Fock representation of gravitational boundary modes and the
  discreteness of the area spectrum},'' {\em Annales Henri Poincare} {\bf 18}
  (2017), no.~11, 3695--3717, \href{http://arXiv.org/abs/1706.00479}{{\tt
  arXiv:1706.00479}}.

\bibitem{CarrozzaHoehn21}
S.~Carrozza and P.~A. Höhn, ``{Edge modes as reference frames and boundary
  actions from post-selection},'' {\em JHEP} {\bf 02} (2022) 172,
  \href{http://arXiv.org/abs/2109.06184}{{\tt arXiv:2109.06184}}.

\bibitem{CEH22}
S.~Carrozza, S.~Eccles, and P.~A. Höhn, ``{Edge modes as dynamical frames:
  charges from post-selection in generally covariant theories},'' May, 2022.
\newblock \href{https://arxiv.org/abs/2205.00913}{arXiv:2205.00913}.

\bibitem{GHK22}
C.~Goeller, P.~A. Höhn, and J.~Kirklin, ``{Diffeomorphism-invariant
  observables and dynamical frames in gravity: reconciling bulk locality with
  general covariance},'' Jun., 2022.
\newblock \href{https://arxiv.org/abs/2206.01193}{arXiv:2206.01193}.

\bibitem{loveridge_symmetry_2018}
L.~Loveridge, T.~Miyadera, and P.~Busch, ``Symmetry, {Reference} {Frames}, and
  {Relational} {Quantities} in {Quantum} {Mechanics},'' {\em Foundations of
  Physics} {\bf 48} (Feb., 2018) 135--198,
  \href{http://arXiv.org/abs/1703.10434}{{\tt arXiv:1703.10434}}.

\bibitem{Giacomini_2019}
F.~{Giacomini}, E.~{Castro-Ruiz}, and {\v{C}}.~{Brukner}, ``{Quantum mechanics
  and the covariance of physical laws in quantum reference frames},'' {\em
  Nature Communications} {\bf 10} (Jan., 2019) 494,
  \href{http://arXiv.org/abs/1712.07207}{{\tt arXiv:1712.07207}}.

\bibitem{Giacomini_spin}
F.~Giacomini, E.~Castro-Ruiz, and {\v{C}}.~Brukner, ``Relativistic Quantum
  Reference Frames: The Operational Meaning of Spin,'' {\em Physical Review
  Letters} {\bf 123} (aug, 2019) \href{http://arXiv.org/abs/1811.08228}{{\tt
  arXiv:1811.08228}}.

\bibitem{vanrietvelde2018change}
A.~Vanrietvelde, P.~A. Höhn, F.~Giacomini, and E.~Castro-Ruiz, ``A change of
  perspective: switching quantum reference frames via a perspective-neutral
  framework,'' {\em Quantum} {\bf 4} (Jan, 2020) 225,
  \href{http://arXiv.org/abs/1809.00556}{{\tt arXiv:1809.00556}}.

\bibitem{delaHamette2020}
A.-C. de~la Hamette and T.~D. Galley, ``Quantum reference frames for general
  symmetry groups,'' {\em {Quantum}} {\bf 4} (Nov., 2020) 367,
  \href{http://arXiv.org/abs/2004.14292}{{\tt arXiv:2004.14292}}.

\bibitem{hoehn2019trinity}
P.~A. Höhn, A.~R. Smith, and M.~P. Lock, ``Trinity of relational quantum
  dynamics,'' {\em Physical Review D} {\bf 104} (Sep, 2021)
  \href{http://arXiv.org/abs/arXiv:1912.00033}{{\tt arXiv:arXiv:1912.00033}}.

\bibitem{castroruiz2021relative}
E.~Castro-Ruiz and O.~Oreshkov, ``Relative subsystems and quantum reference
  frame transformations,'' 2021.
\newblock \href{https://arxiv.org/abs/2110.13199}{arXiv:2110.13199}.

\bibitem{delahamette2021perspectiveneutral}
A.-C. de~la Hamette, T.~D. Galley, P.~A. Höhn, L.~Loveridge, and M.~P.
  Müller, ``Perspective-neutral approach to quantum frame covariance for
  general symmetry groups,'' 2021.
\newblock \href{https://arxiv.org/abs/2110.13824}{arXiv:2110.13824}.

\bibitem{Carlip:1996yb}
S.~Carlip, ``{The Statistical mechanics of the three-dimensional Euclidean
  black hole},'' {\em Phys. Rev. D} {\bf 55} (1997) 878--882,
\href{http://arXiv.org/abs/gr-qc/9606043}{{\tt arXiv:gr-qc/9606043}}.
%%CITATION = GR-QC/9606043;%%.

\bibitem{HarlowWu20}
D.~{Harlow} and J.-q. {Wu}, ``{Covariant phase space with boundaries},'' {\em
  Journal of High Energy Physics} {\bf 2020} (Oct., 2020) 146,
  \href{http://arXiv.org/abs/1906.08616}{{\tt arXiv:1906.08616}}.

\bibitem{Sachs}
R.~Sachs, ``Asymptotic Symmetries in Gravitational Theory,'' {\em Phys. Rev.}
  {\bf 128} (Dec, 1962) 2851--2864.

\bibitem{Bondi21}
H.~Bondi, M.~G.~J. van~der Burg, and A.~W.~K. Metzner, ``Gravitational waves in
  general relativity, VII. Waves from axi-symmetric isolated system,'' {\em
  Proc. of the Royal Soc. Lond. A: Mathematical, Physical and Engineering
  Sciences} {\bf 269} (1962), no.~1336, 21--52.

\bibitem{AshtekarNullInfinity}
A.~Ashtekar, {\em {Asymptotic Quantization}}.
\newblock Bibliopolis, Napoli, 1987.
\newblock Based on 1984 Naples Lectures.

\bibitem{Ashtekar:1978zz}
A.~Ashtekar and R.~O. Hansen, ``{A unified treatment of null and spatial
  infinity in general relativity. I - Universal structure, asymptotic
  symmetries, and conserved quantities at spatial infinity},'' {\em J. Math.
  Phys.} {\bf 19} (1978)
1542--1566.
%%CITATION = JMAPA,19,1542;%%.

\bibitem{Balachandran:1994up}
A.~P. Balachandran, L.~Chandar, and A.~Momen, ``{Edge states in gravity and
  black hole physics},'' {\em Nucl. Phys.} {\bf B461} (1996) 581--596,
\href{http://arXiv.org/abs/gr-qc/9412019}{{\tt arXiv:gr-qc/9412019}}.
%%CITATION = GR-QC/9412019;%%.

\bibitem{Carlip95}
S.~Carlip, ``Statistical mechanics of the (2+1)-dimensional black hole,'' {\em
  Phys. Rev. D} {\bf 51} (Jan, 1995) 632--637,
  \href{http://arXiv.org/abs/gr-qc/9409052}{{\tt arXiv:gr-qc/9409052}}.

\bibitem{IyerWald94}
V.~Iyer and R.~M. Wald, ``{Some properties of Noether charge and a proposal for
  dynamical black hole entropy},'' {\em Physical Review D} {\bf 50} (1994)
  846--864, \href{http://arXiv.org/abs/gr-qc/9403028}{{\tt
  arXiv:gr-qc/9403028}}.

\bibitem{Ciambelli:2022vot}
L.~Ciambelli, ``{From Asymptotic Symmetries to the Corner Proposal},'' {\em
  PoS} {\bf Modave2022} (2023) 002, \href{http://arXiv.org/abs/2212.13644}{{\tt
  arXiv:2212.13644}}.

\bibitem{Freidel:2023bnj}
L.~Freidel, M.~Geiller, and W.~Wieland, {\em Handbook of Quantum Gravity},
  ch.~{Corner symmetry and quantum geometry}.
\newblock Springer, 2023.
\newblock \href{http://arXiv.org/abs/2302.12799}{{\tt arXiv:2302.12799}}.

\bibitem{Marsden1974}
J.~Marsden and A.~Weinstein, ``Reduction of symplectic manifolds with
  symmetry,'' {\em Reports on Mathematical Physics} {\bf 5} (1974), no.~1,
  121--130.

\bibitem{Poisson:2004ti}
E.~Poisson, ``The Motion of Point Particles in Curved Spacetime,'' {\em Living
  Reviews in Relativity} {\bf 7} (2004), no.~1, 6,
  \href{http://arXiv.org/abs/1102.0529}{{\tt arXiv:1102.0529}}.

\bibitem{Gralla:2008fg}
S.~E. Gralla and R.~M. Wald, ``{A Rigorous Derivation of Gravitational
  Self-force},'' {\em Class. Quant. Grav.} {\bf 25} (2008) 205009,
  \href{http://arXiv.org/abs/0806.3293}{{\tt arXiv:0806.3293}}.

\bibitem{Aichelburg1971}
P.~C. Aichelburg and R.~U. Sexl, ``On the gravitational field of a massless
  particle,'' {\em General Relativity and Gravitation} {\bf 2} (1971), no.~4,
  303--312.

\bibitem{Balasin:1993kf}
H.~Balasin and H.~Nachbagauer, ``{Distributional energy momentum tensor of the
  Kerr-Newman space-time family},'' {\em Class. Quant. Grav.} {\bf 11} (1994)
  1453--1462, \href{http://arXiv.org/abs/gr-qc/9312028}{{\tt
  arXiv:gr-qc/9312028}}.

\bibitem{Gomes:2016mwl}
H.~Gomes and A.~Riello, ``{The observer's ghost: notes on a field space
  connection},'' {\em JHEP} {\bf 05} (2017) 017,
\href{http://arXiv.org/abs/1608.08226}{{\tt arXiv:1608.08226}}.
%%CITATION = ARXIV:1608.08226;%%.

\bibitem{Arnowitt:1962hi}
R.~L. Arnowitt, S.~Deser, and C.~W. Misner, ``{The Dynamics of general
  relativity},'' {\em Gen. Rel. Grav.} {\bf 40} (2008) 1997--2027,
\href{http://arXiv.org/abs/gr-qc/0405109}{{\tt arXiv:gr-qc/0405109}}.
%%CITATION = GR-QC/0405109;%%.

\bibitem{ADMEnergy}
R.~{Arnowitt}, S.~{Deser}, and C.~W. {Misner}, ``{Dynamical Structure and
  Definition of Energy in General Relativity},'' {\em Physical Review} {\bf
  116} (1959) 1322--1330.

\bibitem{ReggeTeitelboim}
T.~Regge and C.~Teitelboim, ``Role of surface integrals in the Hamiltonian
  formulation of general relativity,'' {\em Annals of Physics} {\bf 88} (1974),
  no.~1, 286--318.

\bibitem{AshtekarAsymptcs1980}
A.~{Ashtekar}, ``{Asymptotic Structure of the Gravitational Field at Spatial
  Infinity},'' in {\em General Relativity and Gravitation II}, A.~{Held}, ed.,
  vol.~2, p.~37.
\newblock Plenum Press, New York, 1980.

\bibitem{Beig:1987zz}
R.~Beig and N.~{{\'O}\,Murchadha}, ``{The Poincare group as the symmetry group
  of canonical general relativity},'' {\em Annals Phys.} {\bf 174} (1987)
463--498.
%%CITATION = APNYA,174,463;%%.

\bibitem{thiemann}
T.~Thiemann, {\em Modern Canonical Quantum General Relativity}.
\newblock Cambridge University Press, Cambridge, 2008.

\bibitem{LOSTtheorem}
J.~Lewandowski, A.~Okolow, H.~Sahlmann, and T.~Thiemann, ``{Uniqueness of
  diffeomorphism invariant states on holonomy-flux algebras},'' {\em Commun.
  Math. Phys.} {\bf 267} (2006) 703--733,
\href{http://arXiv.org/abs/gr-qc/0504147}{{\tt arXiv:gr-qc/0504147}}.
%%CITATION = GR-QC/0504147;%%.

\bibitem{Isham92}
C.~J. {Isham}, ``{Canonical Quantum Gravity and the Problem of Time},'' {\em
  arXiv e-prints} (Oct., 1992) gr--qc/9210011,
  \href{http://arXiv.org/abs/gr-qc/9210011}{{\tt arXiv:gr-qc/9210011}}.

\bibitem{Ditt04}
B.~Dittrich, ``{Partial and complete observables for Hamiltonian constrained
  systems},'' {\em Gen. Rel. Grav.} {\bf 39} (2007) 1891--1927,
\href{http://arXiv.org/abs/gr-qc/0411013}{{\tt arXiv:gr-qc/0411013}}.
%%CITATION = GR-QC/0411013;%%.

\bibitem{Giesel:2007wn}
K.~Giesel and T.~Thiemann, ``{Algebraic quantum gravity (AQG). IV. Reduced
  phase space quantisation of loop quantum gravity},'' {\em Class.Quant.Grav.}
  {\bf 27} (2010) 175009,
\href{http://arXiv.org/abs/0711.0119}{{\tt arXiv:0711.0119}}.
%%CITATION = ARXIV:0711.0119;%%.

\bibitem{HennauxTeitelboim}
M.~Henneaux and C.~Teitelboim, {\em Quantization of Gauge Systems}.
\newblock Princeton University Press, Princeton, 1992.

\bibitem{Rovelli_1991}
C.~Rovelli, ``Quantum reference systems,'' {\em Classical and Quantum Gravity}
  {\bf 8} (feb, 1991) 317--331.

\bibitem{Brown:1994py}
J.~D. Brown and K.~V. Kuchar, ``{Dust as a standard of space and time in
  canonical quantum gravity},'' {\em Phys. Rev. D} {\bf 51} (1995) 5600--5629,
  \href{http://arXiv.org/abs/gr-qc/9409001}{{\tt arXiv:gr-qc/9409001}}.

\bibitem{Barnich:2011mi}
G.~Barnich and C.~Troessaert, ``{BMS charge algebra},'' {\em JHEP} {\bf 12}
  (2011) 105,
\href{http://arXiv.org/abs/1106.0213}{{\tt arXiv:1106.0213}}.
%%CITATION = ARXIV:1106.0213;%%.

\bibitem{Speranza:2017gxd}
A.~J. Speranza, ``{Local phase space and edge modes for
  diffeomorphism-invariant theories},'' {\em JHEP} {\bf 02} (2018) 021,
\href{http://arXiv.org/abs/1706.05061}{{\tt arXiv:1706.05061}}.
%%CITATION = ARXIV:1706.05061;%%.

\bibitem{Ciambelli:2022cfr}
L.~Ciambelli and R.~G. Leigh, ``{Universal corner symmetry and the orbit method
  for gravity},'' {\em Nucl. Phys. B} {\bf 986} (2023) 116053,
  \href{http://arXiv.org/abs/2207.06441}{{\tt arXiv:2207.06441}}.

\bibitem{Wieland:2021eth}
W.~Wieland, ``{Barnich\textendash{}Troessaert bracket as a Dirac bracket on the
  covariant phase space},'' {\em Class. Quant. Grav.} {\bf 39} (2022), no.~2,
  025016, \href{http://arXiv.org/abs/2104.08377}{{\tt arXiv:2104.08377}}.

\bibitem{Kabel:2022efn}
V.~Kabel and W.~Wieland, ``{Metriplectic geometry for gravitational
  subsystems},'' {\em Phys. Rev. D} {\bf 106} (2022), no.~6, 064053,
  \href{http://arXiv.org/abs/2206.00029}{{\tt arXiv:2206.00029}}.

\bibitem{rovelli}
C.~Rovelli, {\em Quantum Gravity}.
\newblock Cambridge University Press, Cambridge, 2008.

\bibitem{alexreview}
A.~Perez, ``{The Spin-Foam Approach to Quantum Gravity},'' {\em Living Rev.
  Rel.} {\bf 16} (2013), no.~3,
\href{http://arXiv.org/abs/1205.2019}{{\tt arXiv:1205.2019}}.
%%CITATION = ARXIV:1205.2019;%%.

\bibitem{lqcspinfoam1}
A.~Ashtekar, M.~Campiglia, and A.~Henderson, ``{Casting Loop Quantum Cosmology
  in the Spin Foam Paradigm},'' {\em Class. Quant. Grav.} {\bf 27} (2010)
  135020, \href{http://arXiv.org/abs/1001.5147}{{\tt arXiv:1001.5147}}.

\bibitem{Thiemann:2013lka}
T.~Thiemann and A.~Zipfel, ``{Linking covariant and canonical LQG II: Spin foam
  projector},'' {\em Class. Quant. Grav.} {\bf 31} (2014) 125008,
  \href{http://arXiv.org/abs/1307.5885}{{\tt arXiv:1307.5885}}.

\bibitem{Apadula2022}
L.~Apadula, E.~Castro-Ruiz, and {\v{C}}.~Brukner, ``{Quantum Reference Frames
  for Lorentz Symmetry},'' 12, 2022.
\newblock {\href{https://arxiv.org/abs/2212.14081}{arXiv:2212.14081}}.

\bibitem{delaHamette2022}
A.-C. de~la Hamette, V.~Kabel, M.~Christodoulou, and {\v{C}}.~Brukner,
  ``{Quantum diffeomorphisms cannot make indefinite causal order definite},''
  11, 2022.
\newblock {\href{https://arxiv.org/abs/2211.15685}{arXiv:2211.15685}}.

\bibitem{giacomini2021einsteins}
F.~Giacomini and {\v{C}}.~Brukner, ``Einstein's {E}quivalence principle for
  superpositions of gravitational fields,'' 2021.
\newblock \href{https://arxiv.org/abs/2012.13754}{arXiv:2012.13754}.

\bibitem{giacomini22einsteins}
F.~{Giacomini} and {\v{C}}.~{Brukner}, ``{Quantum superposition of spacetimes
  obeys Einstein's equivalence principle},'' {\em AVS Quantum Science} {\bf 4}
  (Mar., 2022) 015601, \href{http://arXiv.org/abs/2109.01405}{{\tt
  arXiv:2109.01405}}.

\bibitem{delaHamette2021falling}
A.-C. de~la Hamette, V.~Kabel, E.~Castro-Ruiz, and {\v{C}}.~Brukner, ``{Quantum
  reference frames for an indefinite metric},'' {\em Commun. Phys.} {\bf 6}
  (2023), no.~1, 231, \href{http://arXiv.org/abs/2112.11473}{{\tt
  arXiv:2112.11473}}.

\bibitem{Strominger:2013jfa}
A.~Strominger, ``{On BMS Invariance of Gravitational Scattering},'' {\em JHEP}
  {\bf 07} (2014) 152, \href{http://arXiv.org/abs/1312.2229}{{\tt
  arXiv:1312.2229}}.

\bibitem{castroruiz2019time}
E.~{Castro-Ruiz}, F.~{Giacomini}, A.~{Belenchia}, and {\v{C}}.~{Brukner},
  ``{Quantum clocks and the temporal localisability of events in the presence
  of gravitating quantum systems},'' {\em Nature Communications} {\bf 11} (May,
  2020) 2672, \href{http://arXiv.org/abs/1908.10165}{{\tt arXiv:1908.10165}}.

\bibitem{cepollaro2021quantum}
C.~Cepollaro and F.~Giacomini, ``Quantum generalisation of {E}instein's
  {E}quivalence {P}rinciple can be verified with entangled clocks as quantum
  reference frames,'' 2021.
\newblock \href{https://arxiv.org/abs/2112.03303}{arXiv:2112.03303}.

\bibitem{Foo2022}
J.~{Foo}, R.~B. {Mann}, and M.~{Zych}, ``{Schr{\"o}dinger{\textquoteright}s
  black hole cat},'' {\em International Journal of Modern Physics D} {\bf 31}
  (Oct., 2022) 2242016--135, \href{http://arXiv.org/abs/2204.00384}{{\tt
  arXiv:2204.00384}}.

\bibitem{Kabel2022conformal}
V.~Kabel, A.-C. de~la Hamette, E.~Castro-Ruiz, and {\v{C}}.~Brukner, ``Quantum
  conformal symmetries for spacetimes in superposition,'' 2022.
\newblock \href{https://arxiv.org/abs/2207.00021}{arXiv:2207.00021}.

\bibitem{Oriti09}
D.~{Oriti}, {\em {Approaches to Quantum Gravity}}.
\newblock Cambridge University Press, 2009.

\bibitem{lollreview}
R.~Loll, ``{Discrete approaches to quantum gravity in four-dimensions},'' {\em
  Living Rev.Rel.} {\bf 1} (1998) 13,
\href{http://arXiv.org/abs/gr-qc/9805049}{{\tt arXiv:gr-qc/9805049}}.
%%CITATION = GR-QC/9805049;%%.

\bibitem{Loll:2019rdj}
R.~Loll, ``{Quantum Gravity from Causal Dynamical Triangulations: A Review},''
  {\em Class. Quant. Grav.} {\bf 37} (2020), no.~1, 013002,
  \href{http://arXiv.org/abs/1905.08669}{{\tt arXiv:1905.08669}}.

\bibitem{flppdspinfoam}
J.~Engle, R.~Pereira, and C.~Rovelli, ``Flipped spinfoam vertex and loop
  gravity,'' {\em Nucl. Phys. B} {\bf 798} (2008) 251--290,
  \href{http://arXiv.org/abs/0708.1236v1}{{\tt arXiv:0708.1236v1}}.

\bibitem{Freidel:1998pt}
L.~Freidel and K.~Krasnov, ``{Spin foam models and the classical action
  principle},'' {\em Adv. Theor. Math. Phys.} {\bf 2} (1999) 1183--1247,
\href{http://arXiv.org/abs/hep-th/9807092}{{\tt arXiv:hep-th/9807092}}.
%%CITATION = HEP-TH/9807092;%%.

\bibitem{contphas}
L.~Freidel, M.~Geiller, and J.~Ziprick, ``{Continuous formulation of the Loop
  Quantum Gravity phase space},'' {\em Class. Quant. Grav.} {\bf 30} (2013)
  085013,
\href{http://arXiv.org/abs/1110.4833}{{\tt arXiv:1110.4833}}.
%%CITATION = ARXIV:1110.4833;%%.

\bibitem{Bianchi:2009tj}
E.~Bianchi, ``{Loop Quantum Gravity \`a la Aharonov-Bohm},'' {\em Gen. Rel.
  Grav.} {\bf 46} (2014) 1668,
\href{http://arXiv.org/abs/0907.4388}{{\tt arXiv:0907.4388}}.
%%CITATION = ARXIV:0907.4388;%%.

\bibitem{Wieland:2016exy}
W.~Wieland, ``{Discrete gravity as a topological field theory with light-like
  curvature defects},'' {\em JHEP} {\bf 05} (2017) 142,
  \href{http://arXiv.org/abs/1611.02784}{{\tt arXiv:1611.02784}}.

\bibitem{Wieland:2013cr}
W.~M. Wieland, ``{Hamiltonian spinfoam gravity},'' {\em Class. Quant. Grav.}
  {\bf 31} (2014) 025002, \href{http://arXiv.org/abs/1301.5859}{{\tt
  arXiv:1301.5859}}.

\bibitem{Oriti14}
D.~{Oriti}, ``{Disappearance and emergence of space and time in quantum
  gravity},'' {\em Studies in the History and Philosophy of Modern Physics}
  {\bf 46} (May, 2014) 186--199, \href{http://arXiv.org/abs/1302.2849}{{\tt
  arXiv:1302.2849}}.

\bibitem{Dittrich14}
B.~Dittrich, ``{The continuum limit of loop quantum gravity - a framework for
  solving the theory},'' in {\em Loop Quantum Gravity, The First Thirty Years},
  A.~Abhay and J.~Pullin, eds., vol.~4.
\newblock World Scientific, 2017.
\newblock
\href{http://arXiv.org/abs/1409.1450}{{\tt arXiv:1409.1450}}.
\newblock
%%CITATION = ARXIV:1409.1450;%%.

\bibitem{Asante:2022dnj}
S.~K. Asante, B.~Dittrich, and S.~Steinhaus, ``{Spin foams, Refinement limit
  and Renormalization},'' \href{http://arXiv.org/abs/2211.09578}{{\tt
  arXiv:2211.09578}}.

\end{thebibliography}\endgroup
\bibliographystyle{bibstyle.bst}

\appendix

\section{On the radiative symplectic structure}\label{appdxB}
\noindent In this appendix, we collect the missing details that lead from Eq.~\eref{Omradintermediate} to the expression of the presymplectic two-form of the radiation modes in which the invariance under diffeomorphisms is manifest, i.e.,\ Eq.~\eref{Omrad2}. First of all, we insert the definition of the effective energy-momentum current of the radiative modes. Going back to \eref{Tdef1}, we find
\begin{align}\nonumber
\Omega_{\mathcal{D}}^{\mtext{rad}}  = &\int_{\mathcal{D}}\left[{\ast}\left(\di \big(\bbvar{d}X_{[\mu}\big)\wedge\nT{1}{f}{_{\nu]}}\right)\wedge\bbvar{d}\nT{1}{\Delta}{^{\mu\nu}}+{\ast}\left(\di X_{[\mu}\wedge\bbvar{d}\nT{1}{f}{_{\nu]}}\right)\wedge\bbvar{d}\nT{1}{\Delta}{^{\mu\nu}}+\bbvar{d}{X}_\mu\,\bbvar{d}\nT{2}{t}{^\mu}\right]=\nonumber\\
=&\int_{\mathcal{D}}\biggl[{\ast}\left(\di \big(\bbvar{d}X_{[\mu}\big)\wedge\nT{1}{f}{_{\nu]}}\right)\wedge\bbvar{d}\nT{1}{\Delta}{^{\mu\nu}}+{\ast}\left(\di X_{[\mu}\wedge\bbvar{d}\nT{1}{f}{_{\nu]}}\right)\wedge\bbvar{d}\nT{1}{\Delta}{^{\mu\nu}}+\nonumber\\
&-{\ast}\left(\bbvar{d}X_{[\mu}\nT{1}{f}{_{\nu]}}\right)\wedge\di\big(\bbvar{d}\nT{1}{\Delta}{^{\mu\nu}}\big)-{\ast}\left(\bbvar{d}X_{[\mu}\,\bbvar{d}\nT{1}{f}{_{\nu]}}\right)\wedge\di\nT{1}{\Delta}{^{\mu\nu}}+\nonumber\\
&-{\ast}\left(\bbvar{d}X_{[\mu}\di X_{\nu]}\right)\wedge\big[\nT{1}{\Delta}{},\bbvar{d}\nT{1}{\Delta}{}\big]^{\mu\nu}-\frac{1}{2}{\ast}\left(\bbvar{d}X_{[\mu}\di \big(\bbvar{d}X_{\nu]}\big)\right)\wedge\big[\nT{1}{\Delta}{},\nT{1}{\Delta}{}\big]^{\mu\nu}\biggr]\label{ThetaRad}.
\end{align}
The first, third, and fifth term after the second equality of \eref{ThetaRad} sign can be reorganized as follows:
\begin{align}\nonumber
\int_{\mathcal{D}}\biggl[&{\ast}\left(\di\big(\bbvar{d}X_{[\mu}\big)\wedge\nT{1}{f}{_{\nu]}}\right)\wedge\bbvar{d}\nT{1}{\Delta}{^{\mu\nu}}-{\ast}\left(\bbvar{d}X_{[\mu}\nT{1}{f}{_{\nu]}}\right)\wedge\di\big(\bbvar{d}\nT{1}{\Delta}{^{\mu\nu}}\big)+\\
&-{\ast}\left(\bbvar{d}X_{[\mu}\di X_{\nu]}\right)\wedge\big[\nT{1}{\Delta}{},\bbvar{d}\nT{1}{\Delta}{}\big]^{\mu\nu}\biggr]=\nonumber\\
=&-\int_{\mathcal{D}}\biggl[{\ast}\left(\bbvar{d}X_{[\mu}\di\nT{1}{f}{_{\nu]}}\right)\wedge\bbvar{d}\nT{1}{\Delta}{^{\mu\nu}}+{\ast}\left(\bbvar{d}X_{[\mu}\di X_{\nu]}\right)\wedge\big[\nT{1}{\Delta}{},\bbvar{d}\nT{1}{\Delta}{}\big]^{\mu\nu}\biggr]=\nonumber\\
=&-\int_{\mathcal{D}}\biggl[{\ast}\left(\bbvar{d}X_{[\mu}\di X^\rho\wedge\nT{1}{\Delta}{_{\nu]\rho}}\right)\wedge\bbvar{d}\nT{1}{\Delta}{^{\mu\nu}}+{\ast}\left(\bbvar{d}X_{[\mu}\di X_{\nu]}\right)\wedge\big[\nT{1}{\Delta}{},\bbvar{d}\nT{1}{\Delta}{}\big]^{\mu\nu}\biggr]=\nonumber\\
=&\int_{\mathcal{D}}{\ast}\left(\nT{1}{\Delta}{_{\rho[\mu}}\bbvar{d}X^\rho\wedge\di X_{\nu]}\right)\wedge\bbvar{d}\nT{1}{\Delta}{^{\mu\nu}}.
\end{align}
Consider now the following identity.
\begin{align}\nonumber
\nT{1}{\Delta}{_{\rho\mu}}\bbvar{d}X^\rho&=-\bbvar{X}\hook\di\big(\nT{1}{f}{_\mu}\big)+\big(\bbvar{X}\hook\nT{1}{\Delta}{_{\rho\mu}}\big)\di X^\rho
=\\
&=-\mathcal{L}_{\bbvar{X}}\nT{1}{f}{_\mu}+\di\big(\bbvar{X}\hook\nT{1}{f}{^\mu}\big)+\big(\bbvar{X}\hook\nT{1}{\Delta}{_{\rho\mu}}\big)\di X^\rho.
\end{align}
Consider also
\begin{align}
\int_{\mathcal{D}}{\ast}&\left(\di X_{[\mu}\wedge\di\big(\bbvar{X}\hook\nT{1}{f}{_{\nu]}}\big)\right)\wedge\bbvar{d}\nT{1}{\Delta}{^{\mu\nu}}=
-\int_{\mathcal{D}}{\ast}\left(\di X_{[\mu}\wedge\big(\bbvar{X}\hook\nT{1}{f}{_{\nu]}}\big)\right)\wedge\bbvar{d}\di\nT{1}{\Delta}{^{\mu\nu}}=\nonumber\\
&=\int_{\mathcal{D}}{\ast}\left(\di\big(\bbvar{d} X_{[\mu}\big)\wedge\big(\bbvar{X}\hook\nT{1}{f}{_{\nu]}}\big)\right)\wedge\di\nT{1}{\Delta}{^{\mu\nu}}
=-\int_{\mathcal{D}}{\ast}\left(\bbvar{d} X_{[\mu}\wedge\di\big(\bbvar{X}\hook\nT{1}{f}{_{\nu]}}\big)\right)\wedge\di\nT{1}{\Delta}{^{\mu\nu}}=\nonumber\\
&=\int_{\mathcal{D}}{\ast}\left(\di X_{[\mu}\wedge\di\big(\bbvar{X}\hook\nT{1}{f}{_{\nu]}}\big)\right)\wedge\big(\bbvar{X}\hook\di\nT{1}{\Delta}{^{\mu\nu}}\big)=\nonumber\\&=
\int_{\mathcal{D}}{\ast}\left(\di X_{[\mu}\wedge\di\big(\bbvar{X}\hook\nT{1}{f}{_{\nu]}}\big)\right)\wedge\mathcal{L}_{\bbvar{X}}\nT{1}{\Delta}{^{\mu\nu}}.
\end{align}
Thus
\begin{align}\nonumber
\int_{\mathcal{D}}{\ast}&\left(\nT{1}{\Delta}{_{\rho[\mu}}\bbvar{d}X^\rho\wedge\di X_{\nu]}\right)\wedge\bbvar{d}\nT{1}{\Delta}{^{\mu\nu}}=
\int_{\mathcal{D}}\biggl[-{\ast}\left(\di X_{[\mu}\wedge\mathcal{L}_{\bbvar{X}}\nT{1}{f}{_{\nu]}}\right)\wedge\bbvar{d}\nT{1}{\Delta}{^{\mu\nu}}+\\
&+{\ast}\left(\di X_{[\mu}\wedge\di\big(\bbvar{X}\hook\nT{1}{f}{_{\nu]}}\big)\right)\wedge\mathcal{L}_{\bbvar{X}}\nT{1}{\Delta}{^{\mu\nu}}+{\ast}\left(\big(\bbvar{X}\hook\nT{1}{\Delta}{_{\rho[\mu}}\big)\di X^\rho\wedge \di X_{\nu]}\big)\right)\wedge\bbvar{d}\nT{1}{\Delta}{^{\mu\nu}}\biggr]=\nonumber\\
=&\int_{\mathcal{D}}\biggl[-{\ast}\left(\di X_{[\mu}\wedge\mathcal{L}_{\bbvar{X}}\nT{1}{f}{_{\nu]}}\right)\wedge\bbvar{d}\nT{1}{\Delta}{^{\mu\nu}}+
{\ast}\left(\di X_{[\mu}\wedge\mathcal{L}_{\bbvar{X}}\nT{1}{f}{_{\nu]}}\right)\wedge\mathcal{L}_{\bbvar{X}}\nT{1}{\Delta}{^{\mu\nu}}+\nonumber\\
&-{\ast}\left(\di X_{[\mu}\wedge\bbvar{X}\hook\big(\di\nT{1}{f}{_{\nu]}}\big)\right)\wedge\mathcal{L}_{\bbvar{X}}\nT{1}{\Delta}{^{\mu\nu}}+{\ast}\left(\big(\bbvar{X}\hook\nT{1}{\Delta}{_{\rho[\mu}}\big)\di X^\rho\wedge \di X_{\nu]}\big)\right)\wedge\bbvar{d}\nT{1}{\Delta}{^{\mu\nu}}\biggr].
\end{align}
Next, we turn to the fourth term after the second equality sign in \eref{ThetaRad}. We obtain
\begin{align}
\int_{\mathcal{D}}{\ast}&\left(\bbvar{d}X_{[\mu}\,\bbvar{d}\nT{1}{f}{_{\nu]}}\right)\wedge \di\nT{1}{\Delta}{^{\mu\nu}}=
\int_{\mathcal{D}}{\ast}\left(\di X_{[\mu}\,\bbvar{d}\nT{1}{f}{_{\nu]}}\right)\wedge \left(\bbvar{X}\hook\di\nT{1}{\Delta}{^{\mu\nu}}\right)=\nonumber\\
&=\int_{\mathcal{D}}\left[{\ast}\left(\di X_{[\mu}\,\bbvar{d}\nT{1}{f}{_{\nu]}}\right)\wedge\mathcal{L}_{\bbvar{X}}\nT{1}{\Delta}{^{\mu\nu}}-
{\ast}\left(\di X_{[\mu}\,\bbvar{d}\big(\di\nT{1}{f}{_{\nu]}}\big)\right)\wedge\bbvar{X}\hook\nT{1}{\Delta}{^{\mu\nu}}
\right].
\end{align}
Going back to \eref{ThetaRad} and collecting the various terms, we obtain
\begin{align}
\Omega_{\mathcal{D}}^{\mtext{rad}}  = & \int_{\mathcal{D}}\biggl[{\ast}\left(\di X_{[\mu}\wedge\bbvar{D}\nT{1}{f}{_{\nu]}}\right)\wedge\bbvar{D}\nT{1}{\Delta}{^{\mu\nu}}-\frac{1}{2}{\ast}\left(\bbvar{d}X_{[\mu}\di \big(\bbvar{d}X_{\nu]}\big)\right)\wedge\big[\nT{1}{\Delta}{},\nT{1}{\Delta}{}\big]^{\mu\nu}+\nonumber\\
&-{\ast}\left(\di X_{[\mu}\wedge\bbvar{X}\hook\big(\di\nT{1}{f}{_{\nu]}}\big)\right)\wedge\mathcal{L}_{\bbvar{X}}\nT{1}{\Delta}{^{\mu\nu}}+{\ast}\left(\big(\bbvar{X}\hook\nT{1}{\Delta}{_{\rho[\mu}}\big)\di X^\rho\wedge \di X_{\nu]}\big)\right)\wedge\bbvar{d}\nT{1}{\Delta}{^{\mu\nu}}+\nonumber\\
&+{\ast}\left(\di X_{[\mu}\,\bbvar{d}\big(\di\nT{1}{f}{_{\nu]}}\big)\right)\wedge\bbvar{X}\hook\nT{1}{\Delta}{^{\mu\nu}}\biggr].
\end{align}
The fourth term satisfies
\begin{equation}
{\ast}\left(\big(\bbvar{X}\hook\nT{1}{\Delta}{_{\rho[\mu}}\big)\di X^\rho\wedge \di X_{\nu]}\right)\wedge\bbvar{d}\nT{1}{\Delta}{^{\mu\nu}}=
{\ast}\left(\big(\bbvar{d}\nT{1}{\Delta}{_{\rho[\mu}}\big)\wedge\di X^\rho\wedge \di X_{\nu]}\right)\bbvar{X}\hook\nT{1}{\Delta}{^{\mu\nu}}.
\end{equation}
Thus,
\begin{align}
\Omega_{\mathcal{D}}^{\mtext{rad}}  = & \int_{\mathcal{D}}\biggl[{\ast}\left(\di X_{[\mu}\wedge\bbvar{D}\nT{1}{f}{_{\nu]}}\right)\wedge\bbvar{D}\nT{1}{\Delta}{^{\mu\nu}}-\frac{1}{2}{\ast}\left(\bbvar{d}X_{[\mu}\di \big(\bbvar{d}X_{\nu]}\big)\right)\wedge\big[\nT{1}{\Delta}{},\nT{1}{\Delta}{}\big]^{\mu\nu}+\nonumber\\
&-{\ast}\left(\di X_{[\mu}\wedge\bbvar{X}\hook\big(\di\nT{1}{f}{_{\nu]}}\big)\right)\wedge\mathcal{L}_{\bbvar{X}}\nT{1}{\Delta}{^{\mu\nu}}+{\ast}\left(\di X_{[\mu}\wedge\nT{1}{\Delta}{_{\nu]\rho}}\wedge\di\bbvar{d}X^\rho\right)\bbvar{X}\hook\nT{1}{\Delta}{^{\mu\nu}}\biggr].\label{omradappendix}
\end{align}
The fourth term can be written as
\begin{align}
\int_{\mathcal{D}}&{\ast}\left(\di X_{[\mu}\wedge\nT{1}{\Delta}{_{\nu]\rho}}\wedge\di\bbvar{d}X^\rho\right)\bbvar{X}\hook\nT{1}{\Delta}{^{\mu\nu}}=
\int_{\mathcal{D}}\biggl[-{\ast}\left(\di X_{[\mu}\wedge\di\nT{1}{\Delta}{_{\nu]\rho}}\,\bbvar{d}X^\rho\right)\bbvar{X}\hook\nT{1}{\Delta}{^{\mu\nu}}+\nonumber\\
&-{\ast}\left(\di X_{[\mu}\wedge\nT{1}{\Delta}{_{\nu]\rho}}\,\bbvar{d}X^\rho\right)\wedge \di\big(\bbvar{X}\hook\nT{1}{\Delta}{^{\mu\nu}}\big)\biggr]=\nonumber\\
=&\int_{\mathcal{D}}\biggl[-{\ast}\left(\di X_{[\mu}\wedge\di\nT{1}{\Delta}{_{\nu]\rho}}\,\bbvar{d}X^\rho\right)\bbvar{X}\hook\nT{1}{\Delta}{^{\mu\nu}}+{\ast}\left(\di X_{[\mu}\wedge\bbvar{X}\hook\big(\di\nT{1}{f}{_{\nu]}}\big)\right)\wedge\di\big(\bbvar{X}\hook\nT{1}{\Delta}{^{\mu\nu}}\big)+\nonumber\\
&+{\ast}\left(\di X_{[\mu}\wedge\big(\bbvar{X}\hook\nT{1}{\Delta}{_{\nu]\rho}}\big)\di X^\rho\right)\wedge\di\big(\bbvar{X}\hook\nT{1}{\Delta}{^{\mu\nu}}\big)\Big]=\nonumber\\
=&\int_{\mathcal{D}}\biggl[-{\ast}\left(\di X_{[\mu}\wedge\di\nT{1}{\Delta}{_{\nu]\rho}}\,\bbvar{d}X^\rho\right)\bbvar{X}\hook\nT{1}{\Delta}{^{\mu\nu}}+{\ast}\left(\di X_{[\mu}\wedge\bbvar{X}\hook\big(\di\nT{1}{f}{_{\nu]}}\big)\right)\wedge\di\big(\bbvar{X}\hook\nT{1}{\Delta}{^{\mu\nu}}\big)+\nonumber\\
&-\frac{1}{2}{\ast}\left(\di X_{[\mu}\di X_{\nu]}\right)\wedge\di\left[\bbvar{X}\hook\nT{1}{\Delta}{},\bbvar{X}\hook\nT{1}{\Delta}{}\right]^{\mu\nu}\Big].
\end{align}
Notice that the last term in this equation is a total derivative and hence vanishes, since $\nT{1}{f}{^\mu}$ has compact support. Inserting this expression back into Eq.~\eqref{omradappendix}, we obtain
\begin{align}
\Omega_{\mathcal{D}}^{\mtext{rad}}  = & \int_{\mathcal{D}}\biggl[{\ast}\left(\di X_{[\mu}\wedge\bbvar{D}\nT{1}{f}{_{\nu]}}\right)\wedge\bbvar{D}\nT{1}{\Delta}{^{\mu\nu}}-\frac{1}{2}{\ast}\left(\bbvar{d}X_{[\mu}\di \big(\bbvar{d}X_{\nu]}\big)\right)\wedge\big[\nT{1}{\Delta}{},\nT{1}{\Delta}{}\big]^{\mu\nu}+\nonumber\\
&-{\ast}\left(\di X_{[\mu}\wedge\bbvar{X}\hook\big(\di\nT{1}{f}{_{\nu]}}\big)\right)\bbvar{X}\hook\big(\di\nT{1}{\Delta}{^{\mu\nu}}\big)-{\ast}\left(\di X_{[\mu}\wedge\di\nT{1}{\Delta}{_{\nu]\rho}}\,\bbvar{d}X^\rho\right)\bbvar{X}\hook\nT{1}{\Delta}{^{\mu\nu}}\biggr].
\end{align}
It is now fairly straightforward to show that the second term cancels against the third and fourth. Consider first the third term. Once again taking into account the field equations at order $n=1$, we obtain
\begin{align}
\int_{\mathcal{D}}&{\ast}\left(\di X_{[\mu}\wedge\bbvar{X}\hook\big(\di\nT{1}{f}{_{\nu]}}\big)\right)\bbvar{X}\hook\big(\di\nT{1}{\Delta}{^{\mu\nu}}\big)=
\int_{\mathcal{D}}{\ast}\left(\bbvar{d}X_{[\mu}\,\bbvar{X}\hook\big(\di\nT{1}{f}{_{\nu]}}\big)\right)\wedge \di\nT{1}{\Delta}{^{\mu\nu}}=\nonumber\\
&=\int_{\mathcal{D}}\biggl[-{\ast}\left(\bbvar{d} X_{[\mu}\,\big(\bbvar{X}\hook\nT{1}{\Delta}{_{\nu]\rho}}\big)\di X^\rho\right)\wedge\di\nT{1}{\Delta}{^{\mu\nu}}+{\ast}\left(\bbvar{d} X_{[\mu}\nT{1}{\Delta}{_{\nu]\rho}}\bbvar{d} X^\rho\right)\wedge\di\nT{1}{\Delta}{^{\mu\nu}}\biggr]=\nonumber\\
&=\int_{\mathcal{D}}\biggl[-{\ast}\left(\bbvar{d} X_{[\mu}\,\di X_{\nu]}\right)\wedge\left[\di\nT{1}{\Delta}{},\big(\bbvar{X}\hook\nT{1}{\Delta}{}\big)\right]^{\mu\nu}-\frac{1}{2}{\ast}\left(\bbvar{d} X_{[\mu}\bbvar{d} X_{\nu]}\right)\left[\nT{1}{\Delta}{},\di\nT{1}{\Delta}{}\right]^{\mu\nu}\biggr]=\nonumber\\
%&=\int_{\mathcal{D}}\biggl[-{\ast}\left(\di\nT{1}{\Delta}{_{\rho[\mu}}\bbvar{d} X^{\rho}\wedge\di X_{\nu]}\right)\bbvar{X}\hook\nT{1}{\Delta}{^{\mu\nu}}-\frac{1}{2}{\ast}\left(\bbvar{d} X_{[\mu}\bbvar{d} X_{\nu]}\right)\left[\nT{1}{\Delta}{},\di\nT{1}{\Delta}{}\right]^{\mu\nu}\biggr]=\nonumber\\
&=\int_{\mathcal{D}}\biggl[-{\ast}\left(\di X_{[\mu}\wedge\di\nT{1}{\Delta}{_{\nu]\rho}}\bbvar{d} X^{\rho}\right)\bbvar{X}\hook\nT{1}{\Delta}{^{\mu\nu}}-\frac{1}{2}{\ast}\left(\bbvar{d} X_{[\mu}\di\big(\bbvar{d} X_{\nu]}\big)\right)\wedge\left[\nT{1}{\Delta}{},\nT{1}{\Delta}{}\right]^{\mu\nu}\biggr].
\end{align}
Thus,
\begin{align}
\Omega_{\mathcal{D}}^{\mtext{rad}}  = \int_{\mathcal{D}}{\ast}\left(\di X_{[\mu}\wedge\bbvar{D}\nT{1}{f}{_{\nu]}}\right)\wedge\bbvar{D}\nT{1}{\Delta}{^{\mu\nu}}.
\end{align}

\section{Angular momentum conservation}\label{appdxB2}

%:
\noindent In this appendix, we show that the local angular local is conserved. The calculation simplifies by contracting the local angular momentum with an a constant but otherwise arbitrary $\mathfrak{so}(1,3)$ Lie algebra element $\ou{\omega}{\mu}{\nu}:\di\ou{\omega}{\mu}{\nu}=0$. We define the two-form on spacetime
\begin{equation}
J[\omega] =\frac{1}{2}\mathrm{Tr}\big(\omega J\big):=-\big(2P_{[\mu}X_{\mu]}+S_{\mu\nu}\big)\omega^{\mu\nu}.
\end{equation}
Taking into account the Einstein equations \eref{o2EOM1} and torsionless equation \eref{o2EOM2} at order $n=2$ in the region $\bar{\Sigma}=\Sigma-\mathcal{D}$, where $\nT{1}{f}{^\mu}=0$, we obtain
\begin{align}\nonumber
\di J_{\mu\nu}&=2 P_{[\mu}\wedge \di X_{\nu]}-2 {\ast}\big(\di X_{[\mu}\wedge \di \nT{2}{f}{_{\nu]}}\big)=\\
&=2\,\di X_{[\mu}\wedge\di X^\tau\wedge{\ast}\nT{2}{\Delta}{_{\nu]\tau}}+2{\ast}\Big(\di X_{[\mu}\wedge\nT{2}{\Delta}{_{\nu]\tau}}\Big)\wedge\di X^\tau.
\end{align}
Thus,
\begin{equation}
\frac{1}{2}\di J_{\mu\nu}\omega^{\mu\nu} =-\frac{1}{2}\varepsilon_{\nu\tau\rho\sigma}\di X^\mu\wedge\di X^\tau\wedge\nT{2}{\Delta}{^{\rho\sigma}}\ou{\omega}{\nu}{\mu}-\frac{1}{2}\varepsilon_{\mu\nu\rho\sigma}\di X^\rho\wedge\di X^\tau\wedge\nT{2}{\Delta}{^\sigma_\tau}\omega^{\mu\nu}.
\end{equation}
Using the $SO(1,3)$ invariance of the Levi-Civita tensor, i.e.,
\begin{equation}
\forall\ou{\lambda}{\mu}{\nu}\in\mathfrak{so}(1,3):\varepsilon_{\tau[\mu\nu\rho}\ou{\lambda}{\tau}{\nu]}=0,
\end{equation}
we immediately find $\di J[\omega]=0$. Since $\ou{\omega}{\mu}{\nu}\in\mathfrak{so}(1,3)$ is arbitrary, we conclude
\begin{equation}
\di J_{\mu\nu}=0.
\end{equation}

\end{document}